\documentclass[aps, prd, reprint, showpacs, longbibliography]{revtex4-1}
\pdfoutput=1
\usepackage{graphicx}
\usepackage{amsmath}
\usepackage{amssymb}
\usepackage{braket}
\usepackage[utf8]{inputenc}

\newcommand{\eps}{\varepsilon}
\newcommand{\diff}{\mathrm{d}}
\newcommand{\pathd}{\mathcal{D}}
\newcommand{\dd}[2][]{\frac{\mathrm{d} #1}{\mathrm{d} #2}}
\newcommand{\pd}[2][]{\frac{\partial #1}{\partial #2}}
\renewcommand{\vec}{\boldsymbol}
\newcommand*{\defeq}{\mathrel{\vcenter{\baselineskip0.5ex \lineskiplimit0pt
                     \hbox{\scriptsize.}\hbox{\scriptsize.}}}%
                     =}

\newcommand{\so}{\curvearrowright}
\newcommand{\mat}[1]{\boldsymbol{\mathrm{#1}}}
\newcommand{\norm}[1]{\left\lVert#1\right\rVert}
\newcommand{\abs}[1]{\left|#1\right|}
\DeclareMathOperator{\tr}{tr}
\DeclareMathOperator{\erf}{erf}
\DeclareMathOperator{\csch}{csch}

\newcommand{\up}{\mathrm}

\usepackage{xcolor}

\begin{document}

\title{Discrete worldline instantons}

\author{Christian Schneider$^{1}$}

\author{Greger Torgrimsson$^{2,3}$}

\author{Ralf Sch\"utzhold$^{1,4,5}$}

\affiliation{$^1$Fakult\"at f\"ur Physik, Universit\"at Duisburg-Essen, Lotharstrasse 1, 47057 Duisburg, Germany,}

\affiliation{$^2$Theoretisch-Physikalisches Institut, Abbe Center of Photonics,
Friedrich-Schiller Universit\"at Jena, Max-Wien-Platz 1, 07743 Jena, Germany,}

\affiliation{$^3$Helmholtz Institute Jena, Fr\"obelstieg 3, 07743 Jena, Germany,}

\affiliation{$^4$Helmholtz-Zentrum Dresden-Rossendorf, Bautzner Landstraße 400, 01328 Dresden, Germany,}

\affiliation{$^5$Institut f\"ur Theoretische Physik, Technische Universit\"at Dresden, 01062 Dresden, Germany.}

\date{\today}

\begin{abstract}
  The semiclassical approximation of the worldline path integral is a
  powerful tool to study nonperturbative electron-positron pair
  creation in spacetime-dependent background fields. Finding solutions
  of the classical equations of motion, i.e. worldline instantons, is
  possible analytically only in special cases, and a numerical
  treatment is nontrivial as well. We introduce a completely general
  numerical approach based on an approximate evaluation of the
  discretized path integral that easily and robustly gives the full
  semiclassical pair production rate in nontrivial multi-dimensional
  fields, and apply it to some example cases.
\end{abstract}

\pacs{12.20.-m, 11.15.Kc, 02.70.Bf}

\maketitle

\section{Introduction}
An as of today still experimentally unconfirmed prediction of quantum
electrodynamics is that of nonperturbative electron-positron pair
creation in the presence of a strong electric field~\cite{Sauter1931,
  Heisenberg1936, Hund1941}.  Schwinger~\cite{Schwinger1951} gave the
pair production rate per unit volume $\mathcal{P}_{e^+e^-}$ (or more
properly the rate of vacuum decay~\cite{Cohen2008}) in a constant,
homogeneous electric field $E$ in 3+1 dimensions as ($\hbar=c=1$)
\begin{equation}
  \label{eq:constHom}
  \mathcal{P}_{e^+e^-} = \frac{(qE)^2}{4\pi^3}\sum_{n=1}^{\infty}\frac{1}{n^2}
  \exp\left(-n\pi \frac{m^2}{qE}\right),
\end{equation}
where $q$ is the elementary charge and $m$ the mass of the 
electrons and positrons.
The generalization to inhomogeneous and time dependent background
fields is far from straightforward, since this is a nonperturbative
effect (as is visible from the inverse dependence on $q$ and $E$ in
the exponent of~\eqref{eq:constHom}).
Apart from the fundamental interest in this effect as a prototypical
example for a nonperturbative phenomenon in quantum field theory, a
better understanding is also desirable in view of the various
experimental initiatives aimed at reaching ultra-high field strengths
\footnote{For example the \emph{Extreme Light Infrastructure} project
  \protect\url{https://eli-laser.eu/}}.

It is in general difficult to obtain the pair production
probability for multi-dimensional fields. While there has recently
been some progress~\cite{Kohlfurst2016, Aleksandrov2017b,
  Kohlfurst2018, Lv2018, Aleksandrov2018, Kohlfurst2018b} in direct
numerical computation of the exact probability for multi-dimensional
fields, we will instead focus on an approach using the worldline
path integral.
  
This formulation is an alternative to path integrals
over fields to express amplitudes in quantum field theories. The first
steps in this direction were pioneered by Fock, who expressed
solutions of the Dirac equation via a four dimensional
Schr\"odinger-type equation with space and time parameterized by an
additional parameter~\cite{Fock1937}. After Nambu emphasized how
beneficial this representation would be in the path integral
approach~\cite{Nambu1950}, Feynman derived the Klein-Gordon
propagator~\cite{Feynman1950} and Dirac propagator~\cite{Feynman1951}
in this worldline formulation. In parallel, Schwinger's famous
paper~\cite{Schwinger1951} used a similar representation.

It is possible to approximate this worldline path integral
for inhomogeneous fields numerically using discretization and Monte
Carlo methods~\cite{Gies2001, Gies2003, Gies2005,
  Gies2011}. Although our method is based on discretization as well,
we use an instanton approach to compute the integrals instead of
statistical sampling.

Both Feynman and Schwinger mentioned the four dimensional particle's
equations of motion in the classical limit, but the first explicit
mention of an instanton approximation to the (Euclidean) worldline
path integral was given by Affleck, Alvarez and
Manton~\cite{Affleck1982}. They derived the pair production rate for a
constant homogeneous background field in a way that is very similar to
the method used today. The approach was extended to inhomogeneous
fields, including the sub-leading fluctuation
prefactor~\cite{Dunne2005,Dunne2006a}.

An exact analytic treatment is possible in some simple cases
\cite{Dunne2005,Dunne2006a, Ilderton2015} and analytic approximations
allows us to study suitable limiting cases \cite{Schutzhold2008,
  Gies2016, Schneider2016}, but in general solutions of the instanton
equations of motion have to be found numerically.
This can be done using, e.g., shooting
methods~\cite{Dunne2006}, but the highly nonlinear nature of the
equations of motion makes this approach very unstable.

After briefly sketching the semiclassical approximation of the
worldline path integral in section~\ref{sec:wli-method}, we introduce
a different approach to numerically evaluate the path integral by
discretization in sections~\ref{sec:discretization}
and~\ref{sec:prefactor}, and a method to trace families of solutions
over a range of field parameters in
section~\ref{sec:continuation}. 
Finally, we will apply the method to some example cases, 
both with results known analytically (to assess
the accuracy of the approximation) and new examples to demonstrate 
the scope of the approach in section~\ref{sec:applications}.

\section{Worldline instanton method}
\label{sec:wli-method}

The central object of the method is the effective action
$\Gamma_\up{M}$, defined using the vacuum persistence amplitude
\begin{equation}
  e^{i\Gamma_\up{M}} := \braket{0_\up{out}|0_\up{in}}.
\end{equation}
We take the probability for pair creation to be the complement of the
vacuum remaining stable, so
\begin{equation}
  P^{e^+e^-} = 1 - \left|\braket{0_\up{out}|0_\up{in}}\right|^2 = 1-\left|e^{i\Gamma_\up{M}}\right|^2
  \approx 2 \Im \Gamma_\up{M},
\end{equation}
the subscript $_\up{M}$ denoting the physical, Minkowskian
quantity. We will work with the Euclidean effective action $\Gamma$,
related to the Minkowski expression by $\Gamma_\up{M} = i\Gamma$, so
$\Im\Gamma_\up{M} = \Re\Gamma$~\cite{Dunne2006a}.

The Euclidean worldline path integral for spinor QED reads (see, e.g.,
\cite{Schubert2001, Schubert2012})
\begin{align}
  \label{eq:WL-path-integral}
  \Gamma = &\int_0^\infty\frac{\diff{T}}{T} e^{-\frac{m^2}{2} T}
  \int_{x(T)=x(0)}\pathd x(\tau) \nonumber\\
  &\quad\times \Phi[x] \exp\left(-\int_0^T\diff{\tau}
      \left(\frac{\dot{x}^2}{2} + iq A(x)\cdot \dot{x}
    \right)\right),
\end{align}
where $A_\mu$ is the Euclidean potential representing the 
external electromagnetic field $F_{\mu\nu}$
and $x_\mu(\tau)$ are periodic 
worldlines in Euclidean space parametrized by the ``proper time''
$\tau$ with $\dot x^\mu=\up{d}x^\mu/\up{d}\tau$. 
There exist a couple of different representations of the spin factor,
see \cite{Schubert2001, Gies2005a}. We will use
\begin{equation}
  \Phi[x] = \frac{1}{2}\tr\mathcal{P} e^{\frac{i}{4}\int_0^T\diff{\tau}\sigma_{\mu\nu}qF_{\mu\nu}(x)},
\end{equation}
with $\mathcal{P}$ denoting path ordering, $\tr$ the trace over spinor
indices and $\sigma_{\mu\nu}$ the commutator of the Dirac matrices
\begin{equation}
  \sigma_{\mu\nu} = \frac{1}{2}\left[\gamma_\mu, \gamma_\nu\right].
\end{equation}
For simple fields, the Euclidean potential $A_\mu$
and field tensor $F_{\mu\nu}$ are purely imaginary, so 
$iA_\mu$ and $iF_{\mu\nu}$ are real.

We immediately introduce dimensionless quantities
using some reference field strength $E$, which makes a numerical
treatment possible and simplifies the following derivation,
\begin{equation}
  \tilde{x}_\mu = x_\mu \frac{qE}{m}, \quad
  \tilde{F}_{\mu\nu} = F_{\mu\nu}\frac{1}{E}, \quad
  \tilde{A}_\mu = \frac{qE}{m}\frac{1}{E}A_\mu,
\end{equation}
and also rescale the integration variable
\begin{equation}
  \tilde{T} = qE T, \quad u = \frac{1}{T}\tau = \frac{qE}{\tilde{T}} \tau,
  \quad \so \pd{u} = \frac{\tilde{T}}{qE}\pd{\tau}.
\end{equation}
We can now exchange the order of integration,
\begin{align}
  \label{eq:worldline:pathIntegralScaled}
  \Gamma=
  &\int_{x(1)=x(0)}\pathd{x}(u) 
  \int_0^\infty\diff{\tilde{T}}\,\frac{\Phi[\tilde{x}]}{\tilde{T}}\\
  &\quad\times
  \exp\left(-\frac{m^2}{qE}\left(
      \frac{\tilde{T}}{2}
      + \int\limits_0^{1}\diff{u} \left(
        \frac{\dot{\tilde{x}}^2}{2\tilde{T}}
        + i \dot{\tilde{x}}\cdot \tilde{A}
      \right)
    \right)\right),\nonumber
\end{align}
so we can perform the $\tilde{T}$-integration using Laplace's
method. Due to our rescaling, $m^2/qE$ is singled out as the large
parameter of the expansion, while all other quantities are of order
unity. We obtain the saddle point
$\tilde{T}_0 = \sqrt{\int_0^1\diff{u}\ \dot{\tilde{x}}^2} =:
a[\tilde{x}]$, so including the quadratic fluctuation around the
saddle we arrive at the approximation
\begin{equation}
  \label{eq:WL-T-int}
  \Gamma \approx \int_{x(T)=x(0)}\pathd{x}(\tau)
  \sqrt{\frac{2\pi}{a[\tilde{x}]}\frac{qE}{m^2}}
  \Phi[\tilde{x}] e^{-\frac{m^2}{qE}\mathcal{A}[\tilde{x}]},
\end{equation}
with the non-local (due to $a[\tilde{x}]$) action
\begin{equation}
  \mathcal{A}[\tilde{x}] = a[\tilde{x}]
  + \int\limits_0^1\diff{u}\ \dot{\tilde{x}}\cdot i\tilde{A}(\tilde{x}),
\end{equation}
and the spin factor
\begin{equation}
  \Phi[\tilde{x}] = \frac{1}{2}\tr\mathcal{P}\exp\left({\frac{a[\tilde{x}]}{4}
      \int_0^1\diff{u}\ \sigma_{\mu\nu}i\tilde{F}_{\mu\nu}(\tilde{x})}\right).
\end{equation}

Note that in \eqref{eq:WL-T-int} we symbolically restored the original
parametrization $\pathd{x}(\tau)$ in the path integral differential,
this will be relevant for the discretization in the next section.

Applying Laplace's method to the path integral, we need to find a path
$\tilde{x}_\mu(u)$ that satisfies the periodic boundary conditions and
extremizes the exponent in~\eqref{eq:WL-T-int}, so a solution to the
Euler-Lagrange equations (the Lorentz force equation in this case)
\begin{equation}
  \label{eq:instanton-equations-sqrt}
  \frac{\ddot{\tilde{x}}_\mu}{a[\tilde{x}]} = iq\tilde{F}_{\mu\nu}\dot{\tilde{x}}_\nu.
\end{equation}
Contracting~\eqref{eq:instanton-equations-sqrt} with
$\dot{\tilde{x}}_\mu$ we see that (due to the antisymmetry of
$\tilde{F}_{\mu\nu}$) $\dot{\tilde{x}}^2 = \text{const.} = a^2$,
simplifying the instanton equations of motion to
\begin{equation}
  \label{eq:instanton-equations}
  \ddot{\tilde{x}}_\mu = iaq\tilde{F}_{\mu\nu}\dot{\tilde{x}}_\nu.
\end{equation}
The prefactor of the Laplace approximation is given by the second
variation of the action around the classical solution
to~\eqref{eq:instanton-equations}, amounting to an operator
determinant. The determinant has to be defined carefully, but we can
completely sidestep this complication by instead performing Laplace's
method \emph{after} discretization, when we can calculate the
fluctuation prefactor by standard methods of linear algebra.

\section{Discretization}
\label{sec:discretization}

We approximate~\eqref{eq:WL-T-int} by discretizing the trajectories
$\tilde{x}_\mu(u)$ into $N$ $d$-dimensional points (in general $d = 3+1$, but
for simple field configurations it is possible to only consider
$d=1+1$ or $d=2+1$ dimensions, so we will keep the dimensionality variable):
\begin{equation}
  \label{eq:trajectory-discretization}
  \tilde{x}_\mu^i \defeq \tilde{x}_\mu\left(\frac{l}{N}\right), \quad l = 0, 1, \dots, N-1.
\end{equation}
The velocity is then approximated using (forward) finite differences
\begin{equation}
  \label{eq:velocity-discretization}
  \dot{\tilde{x}}_\mu^l \approx \frac{\tilde{x}_\mu^{l+1} - \tilde{x}_\mu^l}{\eps}, \quad\eps = \frac{1}{N},
\end{equation}
with the identification $\tilde{x}_\mu^N\equiv \tilde{x}_\mu^0$.

Discretizing the path integral requires a normalization factor for
each $\tilde{x}_\mu^i$-integral. We could find these factors by
performing the integration in the free case, however this is not
necessary: In the derivation of \eqref{eq:WL-path-integral} the path
integral arises as an ordinary nonrelativistic transition amplitude,
so we can use Feynman's normalization $1/A$ for each integral (see,
e.g., \cite{feynman1948}), with
\begin{equation}
  A = \sqrt{\frac{2\pi T_0}{N}} = \sqrt{\frac{2\pi a[x]}{qE N}}.
\end{equation}
Using this normalization and replacing the $N\times d$ integrations by
the dimensionless versions we arrive at the discretized worldline path
integral
\begin{align}
  \label{eq:WL-discretized-N}
  \Gamma \approx &\left(\prod_{k=0}^{N-1}\int\diff{^d\tilde{x}^k}\right)
  \left(\frac{m^2}{qE} \frac{N}{2\pi a[\tilde{x}_\mu^l]}\right)^{Nd/2} \nonumber\\
  & \qquad\times\sqrt{\frac{2\pi}{a[\tilde{x}_\mu^l]}\frac{qE}{m^2}}
           \Phi[\tilde{x}_\mu^l] e^{-\frac{m^2}{qE}\mathcal{A}[\tilde{x}_\mu^l]}.
\end{align}
As we have now expressed everything in terms of the dimensionless
variables, we will from now on drop the tilde. We still need
discretized expressions for $a$, $\mathcal{A}$ and $\Phi$,
\begin{align}
  a[x_\mu^l] &\defeq \sqrt{N\sum_{k=0}^{N-1}(x_\nu^{k+1}-x_\nu^k)^2} \\
  \mathcal{A}[x_\mu^l] &\defeq a[x_\mu^l] \nonumber\\
  \label{eq:gaugeDiscretization}
  & + \sum_{k=0}^{N-1}\left(\frac{A_\nu(x_\mu^{k+1})+A_\nu(x_\mu^k)}{2}\right)
                         (x_\nu^{k+1}-x_\nu^k),
\end{align}
the square brackets denoting dependence on all points, instead of a
particular choice of indices.

The form of discretization of the gauge term is not at all obvious,
other choices like having just $A_\nu(x_\mu^k)$ or
$A_\nu(x_\mu^{k+1})$ or evaluating the gauge field between points
$A_\nu((x_\mu^k+x_\mu^{k+1})/2)$ would yield the same classical
continuum limit. That does not mean however that the resulting
propagator is the same, see~\cite{Rabello1995, Stone2000, Gaveau2004,
  Schulman2005}. The midpoint prescription
in~\eqref{eq:gaugeDiscretization} arises when the path integral
representation is derived from the vacuum persistence amplitude using
the time slicing procedure, see e.g.~\cite{DHoker1996}.

Special care has to be taken to define a discretized expression for
the spin factor that obeys path ordering. Instead of approximating the
integral by summation and taking the exponential, we employ the
product representation of the exponential function which is
automatically path ordered (cf. \cite{Gies2002}),
\begin{equation}
  \Phi[x_\lambda^l] \defeq \tr\left[\prod_{k=0}^{N-1}
    \left(1 + \frac{a[x_\lambda^l]}{4N}\sigma_{\mu\nu}iF_{\mu\nu}(x_\lambda^k)\right)
  \right].
\end{equation}

The finite dimensional integral~\eqref{eq:WL-discretized-N} can now be
approximated using Laplace's method as well, by finding an
$N\times d$-dimensional vector $\bar{x}_\mu^l$ (a \emph{discrete
  worldline instanton}) that extremizes the action function
$\mathcal{A}[x_\mu^l]$, that is
\begin{equation}
  \left.\dd[\mathcal{A}]{x_\mu^l}\right|_{x_\mu^l=\bar{x}_\mu^l} = 0.
\end{equation}

To ease notation, we will condense the proper time index $l$ and the
spacetime index $\mu$ into a single vector
\begin{equation}
  \vec{X} = (x_1^0, x_2^0, \dots, x_d^0, x_1^1, \dots, x_d^{N-1}),
\end{equation}
so a discrete instanton $\vec{\bar{X}}$ has the property
\begin{equation}
  \label{eq:discretized-eom}
  \vec{F(\vec{\bar{X}})}\defeq\left.\nabla \mathcal{A}(\vec{X})\right|_{\vec{\bar{X}}} = \vec{0}.
\end{equation}
Equation~\eqref{eq:discretized-eom} describes a system of $N\times d$
nonlinear equations in $N\times d$ unknowns, which can be solved
numerically using the Newton-Raphson method or a similar root finding
scheme.

In this discretized picture, the fluctuation prefactor is readily
computed as well, via the determinant of the Hessian of $\mathcal{A}$
\begin{equation}
  \mat{H}(\vec{\bar{X}}) = (\nabla\otimes\nabla) \left.\mathcal{A}(\vec{X})\right|_{\vec{\bar{X}}},
\end{equation}
giving the full semiclassical approximation of the discretized
worldline path integral
\begin{equation}
  \label{eq:discrete-unreg}
  \Gamma \approx
           \sqrt{\frac{2\pi}{a^\up{cl}}\frac{qE}{m^2}}
           \left(\frac{N}{a^\up{cl}}\right)^{Nd/2}
           \frac{\Phi[\vec{\bar{X}}]
           e^{-\frac{m^2}{qE}\mathcal{A}[\vec{\bar{X}}]}}{\sqrt{\det{\mat{H}[\vec{\bar{X}}]}}},
\end{equation}
with $a^\up{cl}\defeq a[\vec{\bar{X}}]$. If the function
$\mathcal{A}[\vec{\bar{X}}]$ were entirely well-behaved we would be
done now, we would just need to find solutions
of~\eqref{eq:discretized-eom} and plug them
into~\eqref{eq:discrete-unreg}. The Gaussian integration resulting in
the determinant prefactor however is only defined for positive
definite matrices in the exponent, which our Hessian $\mat{H}$ is not.

\section{Regularization of the prefactor}
\label{sec:prefactor}

We have two problems with the Hessian matrix of the action
$\mathcal{A}$. One is that of negative eigenvalues of $\mat{H}$. The
corresponding direction in the Gaussian integration diverges, and the
integral has to be defined by analytic continuation. A single negative
mode (which is present for a static electric field) thus turns the
determinant negative, and the whole
expression~\eqref{eq:discrete-unreg} imaginary. This could seem
troubling at first, as the pair production is given by the real part
of the Euclidean effective action. For a field not depending on time
we expect a volume factor from the $x_4$-integration though, which has
to be purely imaginary for a real temporal volume factor
$V_t = -i V_{x_4}$.

A more serious technical issue is that of zero modes. One or more zero
eigenvalues of $\mat{H}$ immediately spoil our result, so they have to
be removed from the integration in some way. One zero mode that is
always present in the worldline path integral is the one corresponding
to reparametrization. Due to the periodic boundary conditions we can
move every point of the curve along the trajectory without a change in
action. We would thus like to separate the integration in this
direction (resulting in a ``volume factor'' of the periodicity, in our
rescaled expression just unity) from the other integrations.

We will use the Faddeev-Popov method~\cite{Faddeev1967} to perform
this separation. While it is commonly used to remove gauge-equivalent
configurations from a gauge theory path integral, it can be applied to
this simpler scenario as well. We insert a factor of unity into the
path integral in terms of the identity
\begin{equation}
  1 = \frac{1}{w} \int\!\diff{\lambda}\ \delta(\chi(\lambda))\left|\dd{\lambda}\chi(\lambda)\right|,
\end{equation}
where $\chi(\lambda)$ is some function chosen so that $\chi=0$ fixes
the zero mode, $\lambda$ parametrizes the symmetry and $w$ is the
number of times $\chi(\lambda)=0$ occurs over the integration
interval~\cite{Gordon2015}. The idea is now that the
$\lambda$-integration can be performed due to the symmetry of the path
integral, resulting in the desired volume factor and a Dirac delta
that fixes the corresponding mode. This is especially elegant for a
discrete numerical evaluation of the semiclassical approximation, as
we can use an exponential representation of the delta function
\begin{equation}
  \label{eq:delta-exp}
  \delta(\chi) = \lim_{\eps\to 0} \sqrt{\frac{m^2/qE}{\eps}}
  \exp\left(-\frac{\pi}{\eps} \frac{m^2}{qE}\chi^2\right),
\end{equation}
where the Gaussian integration over the zero mode produces a factor of
$\sqrt{\eps}$ canceling the prefactor, enabling us to simply set
$\eps = 1$. We insert the factor of $m^2/qE$ for convenience, so the
action $\mathcal{A}$ in~\eqref{eq:WL-discretized-N} just gets an
additional term $\pi \chi^2$.

To fix the reparametrization mode, we take (cf. \cite{Gordon2015, Gordon2016, ZinnJustin1996})
\begin{equation}
  \chi_u(\lambda_u) = \frac{2}{(a^\up{cl})^2} \int_0^1\!\diff{u}
  \ \dot{x}_\nu^{\up{cl}}(u)  x_\nu(u + \lambda_u),
\end{equation}
which is chosen so that
\begin{equation}
  \frac{1}{w} \left|\chi_u'(0)\right| = \frac{1}{2}\frac{2}{(a^\up{cl})^2}
  \int_0^1\!\diff{u}\ \dot{x}_\nu^{\up{cl}}(u)\dot{x}_\nu(u)
  \overset{x=x^{\up{cl}}}{=} 1,
\end{equation}
at the saddle point. Due to the translation invariance we can set
$\lambda_u=0$ in the integrand so the $\lambda_u$ integration is equal
to one. This means we only need to add the (discretized version of)
$\chi_u(0)$ to the action as in~\eqref{eq:delta-exp}, the second
derivatives to $\mat{H}$ and a factor of $\sqrt{m^2/qE}$
from~\eqref{eq:delta-exp} to the prefactor, and just proceed as if no
zero mode were present.

Other zero eigenvalues appear if the electric background field does
not depend on all spacetime coordinates. They are of course easier to
deal with, we could just omit the corresponding integrals and add a
volume factor $\tilde{L}_\mu$ (the tilde is to stress that this is in
terms of the dimensionless coordinates) per invariant direction
$x_\mu$. We can, however, treat these just as the reparametrization
direction, which simplifies a numerical implementation that supports
arbitrary fields. Choosing $\chi_\mu$ to be the average of $x_\mu$
along the trajectory we obtain the volume $\tilde{L}_\mu$, and again a
factor of $\sqrt{m^2/qE}$.

To summarize, our final expression for the semiclassical approximation
of the effective action is
\begin{multline}
  \label{eq:discrete}
  \Gamma \approx
  \frac{V_{N_0}}{m^{-N_0}} \left(\frac{qE}{m^2}\right)^{\frac{N_0}{2}}
  \sqrt{\frac{2\pi}{a^\up{cl}}}
  \left(\frac{N}{a^\up{cl}}\right)^{\frac{Nd}{2}}\\
  \times\frac{\Phi[\vec{\bar{X}}]
    e^{-\frac{m^2}{qE}\mathcal{A}[\vec{\bar{X}}]}}{\sqrt{\det{\mat{H}[\vec{\bar{X}}]}}},
\end{multline}
where the appropriate terms of $\chi$ and its derivatives have been
added to $\mathcal{A}$ and $\mat{H}$, $N_0$ is the number of invariant
spacetime directions, and $V_{N_0}$ the corresponding volume factor
(with units reinstated). Note that~\eqref{eq:discrete} unambiguously
contains the full prefactor including spin effects for an arbitrary
background field, without having to resort to limiting cases to
determine any normalization constants. In addition, the reference
field strength $E$ enters only in the combination $qE/m^2$ in front of
the action and in the prefactor, which has two advantages. First,
having found an instanton $\vec{\bar{X}}$, we can
evaluate~\eqref{eq:discrete} for arbitrary values of $qE/m^2$ without
any additional computational effort. Secondly, the accuracy of the
discretization does not depend on the field strength, so there are no
numerical instabilities for small $E$.

\begin{figure}
  \includegraphics{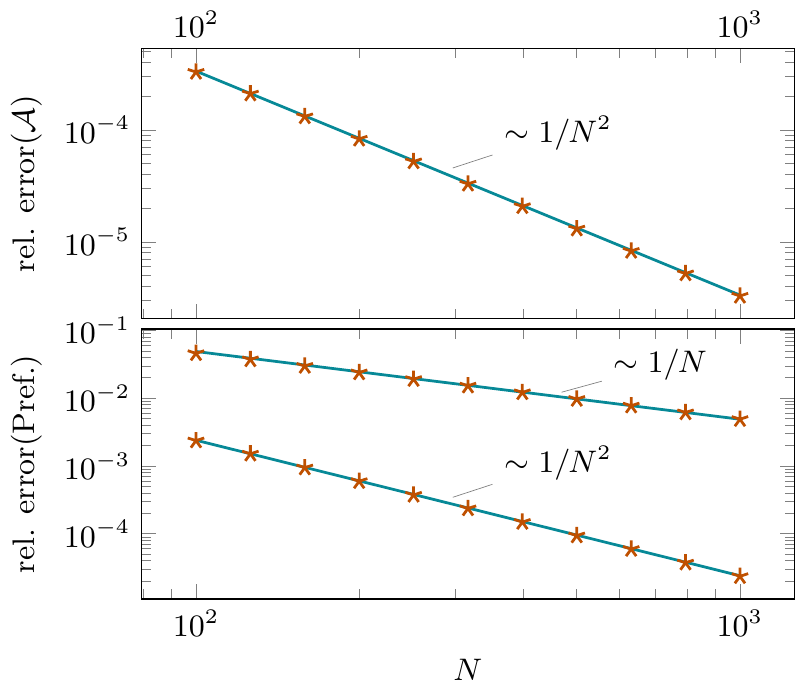}
  \caption{Accuracy of the method for a constant, homogeneous
    field. The error of the prefactor decreases as $1/N$ (the first
    order discretization error) for scalar QED (upper markers in the
    bottom plot), the error of the action as $1/N^2$ (because the
    action has an extremum at that point, upper plot). Interestingly,
    for spinor QED the prefactor decreases as $1/N^2$ as well (lower
    markers).}
  \label{fig:consthom}
\end{figure}

Figure~\ref{fig:consthom} shows how the discretization error scales
with the number of points $N$ for a constant, homogeneous electric
field. For scalar QED (that is, without the spin factor $\Phi$) the
error in the prefactor decreases as $N^{-1}$ as expected for a first
order discretization procedure. As the first variation of the action
vanishes for an instanton, the error of the exponent even decreases as
$N^{-2}$. For spinor QED, on the other hand, the error in the
prefactor decreases as $N^{-2}$ as well. The reason for this is not
obvious, as the only difference is an additional, seemingly
independent multiplicative spin factor.

\section{Numerical continuation}
\label{sec:continuation}

For most fields we are interested in, there is one (or multiple)
parameters that we would like to vary, for example the timescale of a
pulsed field or the inhomogeneity of a spatially varying field
configuration. Let us denote such a parameter $\gamma$. In general we
are interested in the full family of instantons
$\vec{\bar{X}}(\gamma)$. Methods to numerically map such a solution
space are known as \emph{numerical continuation}
algorithms~\cite{Allgower2003, Rheinboldt2000}.

If we know an instanton for a particular value $\gamma_i$ of the
parameter (e.g. the limit $\omega\to 0$ for a time-dependent pulse),
we can use it as the starting point for the numerical solution
of~\eqref{eq:discretized-eom} for a parameter value
$\gamma_{i+1}=\gamma_i+\Delta\gamma$, which is the method used
in~\cite{Gould2017}. If we choose a sufficiently small $\Delta\gamma$,
we can expect the root finding procedure to quickly converge. This
process is called \emph{natural parameter} continuation, because we
vary a physical parameter of the problem at hand, instead of
introducing an artificial variable to blend between an easy and our
actual problem (e.g. solving
$0=\vec{G}(\vec{X}, \gamma) \defeq \gamma\vec{F}(\vec{X}) +
(1-\gamma)\vec{F}_0(\vec{X})$).

Natural parameter continuation works well if the solutions
$\vec{\bar{X}}(\gamma)$ depend on the parameter in a smooth and
uniform manner. If, however, the dependence on $\gamma$ varies
strongly, it is difficult to choose appropriate step lengths
$\Delta\gamma$. For some spatially inhomogeneous fields the instantons
even grow infinitely large in some limit
$\gamma\to\gamma^\text{crit}$, so we need to take ever smaller steps
to reach this value. We could, in principle, adaptively adjust the
step length when the root finding for the next parameter value
converges poorly, but there is an easier method of choosing the
increment $\Delta\gamma$:

Natural parameter continuation can be viewed as a
\emph{predictor-corrector} scheme, with the ``zeroth-order'' predictor
step of just taking the last solution as the starting point for the
next parameter, and performing the numerical root finding as a
corrector step. We can find a better prediction by taking the $\gamma$
derivative of~\eqref{eq:discretized-eom}, yielding the
\emph{Davidenko differential equation}~\cite{Davidenko1953}:
\begin{equation}
  \label{eq:davidenko-ode}
  \vec{0} = \dd{\gamma}\vec{F}(\vec{\bar{X}}, \gamma)
  = \mat{H}(\vec{\bar{X}}, \gamma)\cdot\dd[\vec{\bar{X}}]{\gamma}
  + \pd{\gamma}\vec{F}(\vec{\bar{X}}, \gamma)
\end{equation}
and thus, provided that $\mat{H}$ is invertible (which it is by our
regularization scheme),
\begin{equation}
  \label{eq:davidenko-predict}
  \dd[\vec{\bar{X}}]{\gamma} = -\left(\mat{H}(\vec{\bar{X}}, \gamma)\right)^{-1}
  \cdot\left(\pd{\gamma}\vec{F}(\vec{\bar{X}}, \gamma)\right).
\end{equation}
We can now use~\eqref{eq:davidenko-predict} in two ways: first, having
found an instanton $\vec{\bar{X}}_i$ for a parameter value $\gamma_i$,
it tells us in which way the instanton for a slightly different value
of $\gamma$ differs from the current one, so we can use it as a much
improved predictor in our predictor-corrector scheme, i.e.
$\vec{\bar{X}}_{i+1}\approx \vec{\bar{X}}_i + \Delta\gamma\
\mathrm{d}\vec{\bar{X}}/\mathrm{d}\gamma$. In fact, we could directly
integrate~\eqref{eq:davidenko-predict} to obtain all
solutions. Unfortunately, evaluating the Hessian is costly and we can
afford a much larger step size by performing the corrector steps. As a
compromise it is possible to perform multiple steps according
to~\eqref{eq:davidenko-predict} before starting the root finding
routine. Furthermore, we can use the derivative to scale the step
$\Delta\gamma$ by instead specifying a maximum (or mean) difference
between the points of $\vec{\bar{X}}_i$ and the proposed guess for
$\vec{\bar{X}}_{i+1}$, or even a fixed arclength $\Delta s$ of the
solution curve in $\mathbb{R}^{N\times d + 1}$,
\begin{align}
  \Delta s&=\sqrt{(\Delta\gamma\
    \mathrm{d}\vec{\bar{X}}/\mathrm{d}\gamma)^2+(\Delta \gamma)^2} \nonumber\\
  \label{eq:step-length}
  \Leftrightarrow
  \Delta\gamma &= \frac{\Delta s}{\sqrt{(\mathrm{d}\vec{\bar{X}}/\mathrm{d}\gamma)^2 + 1}}.
\end{align}

A situation may be conceivable where it is not possible to parametrize
the solution set as $\vec{\bar{X}}(\gamma)$ at all, because such a
function would not be single-valued or have infinite slope
somewhere. In this case, we can parametrize both the solution and the
parameter $\gamma$ by a new parameter
\begin{align}
  \vec{\bar{Y}}(u) &= (\vec{\bar{X}}(u), \gamma(u))^\intercal \nonumber\\
  \label{eq:pseudo-arclength}
  \Rightarrow 0 &= \dd{u}\vec{\tilde{F}}(\vec{\bar{Y}}) =
                  \mat{\tilde{H}}\cdot \dd[\vec{\bar{Y}}]{u},
\end{align}
where $\mat{\tilde{H}}$ is now an $(N d + 1) \times (N d)$ matrix, so
\eqref{eq:pseudo-arclength} has to be augmented by an additional
condition. This is chosen to be a constraint on the orientation and
the ``velocity'' of the flow
$1 = \norm{\mathrm{d}\vec{\bar{Y}}/\mathrm{d} u}$, so
$\vec{\bar{Y}}(u)$ is parametrized by arclength, hence the name
\emph{pseudo-arclength continuation} (\emph{pseudo} because this is
only approximately true, as we are taking discrete steps). As long as
$\gamma$ is a suitable parameter, this is equivalent to
\eqref{eq:step-length}, which is what we will be using in the
following.

\section{Applications}
\label{sec:applications}

\begin{figure}
  \centering
  \includegraphics{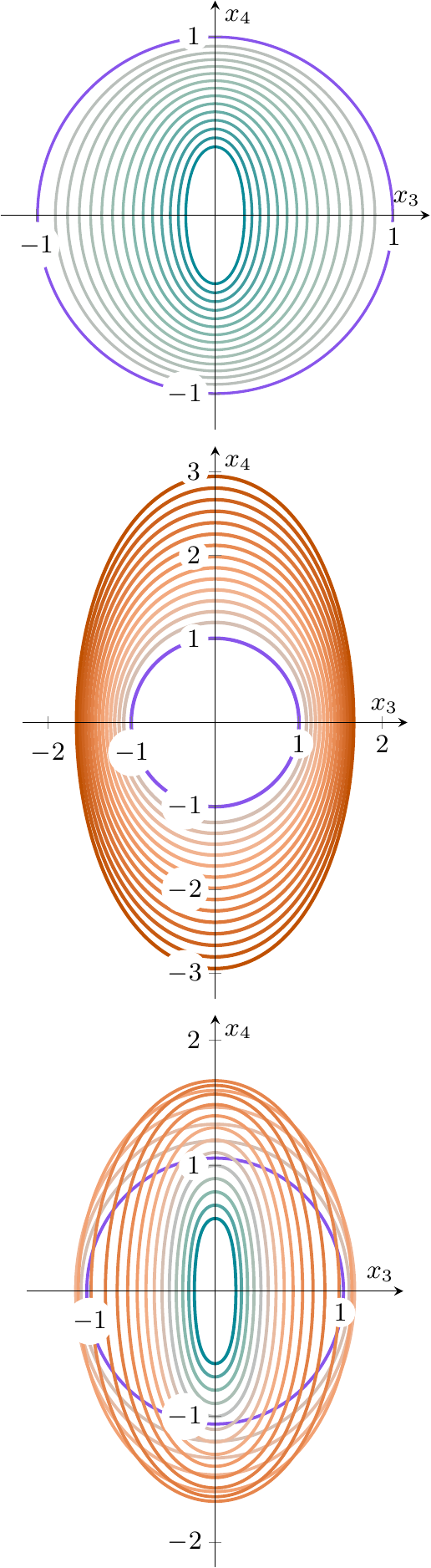}
\caption{Planar instantons for multiple background fields and
  increasing values of $\gamma_{\omega/k}$. Top: temporal Sauter field
  $\vec{E} = E \cosh^{-2}(\omega t) \vec{e}_z$, middle: spatial Sauter
  field $\vec{E} = E \cosh^{-2}(k z) \vec{e}_z$, bottom: spacetime
  bump profile
  $\vec{E} = E \cosh^{-2}(\omega t)\cosh^{-2}(k z) \vec{e}_z$ with
  $k = 3\omega$. The purple trajectories are the limit
  $\gamma_{\omega/k}\to 0$, blue denotes a decrease in action, red an
  increase. As is well known, while temporal variation shrinks the
  instantons and decreases the worldline action (top), spatial
  inhomogeneity has the opposite effect (middle). As the bottom plot
  shows, field configurations are possible that both increase and
  decrease the action in different regimes.}
  \label{fig:onedim}
\end{figure}

\begin{figure}
  \includegraphics{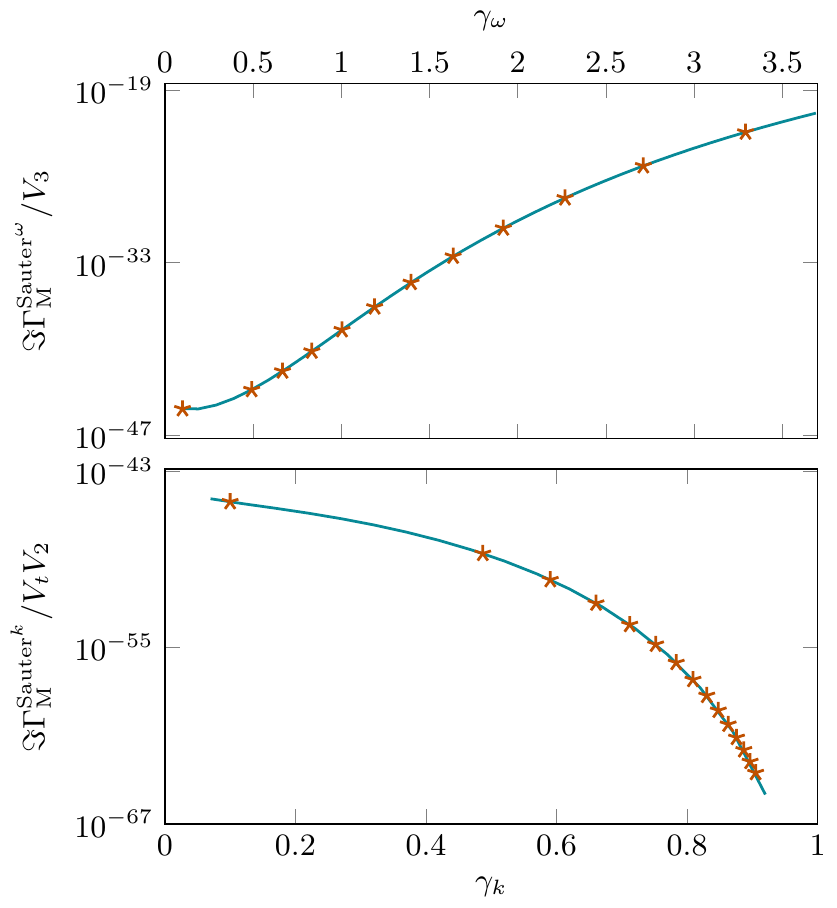}
  \caption{Imaginary part of the Minkowskian effective action (i.e.\
    the pair production rate) for $E=0.033 m^2/q$. Top: temporal
  Sauter pulse, bottom: spatial Sauter profile. Numerical results are
  given by markers and the analytic expressions~\eqref{eq:tSauterAna}
  and~\eqref{eq:xSauterAna} by lines. Note the spacing of markers in
  the spatial case, the step length decreases to keep the overall
  arclength $\Delta s$ constant.}
  \label{fig:sautPP}
\end{figure}

\begin{figure}
  \centering
  \includegraphics[width=\linewidth]{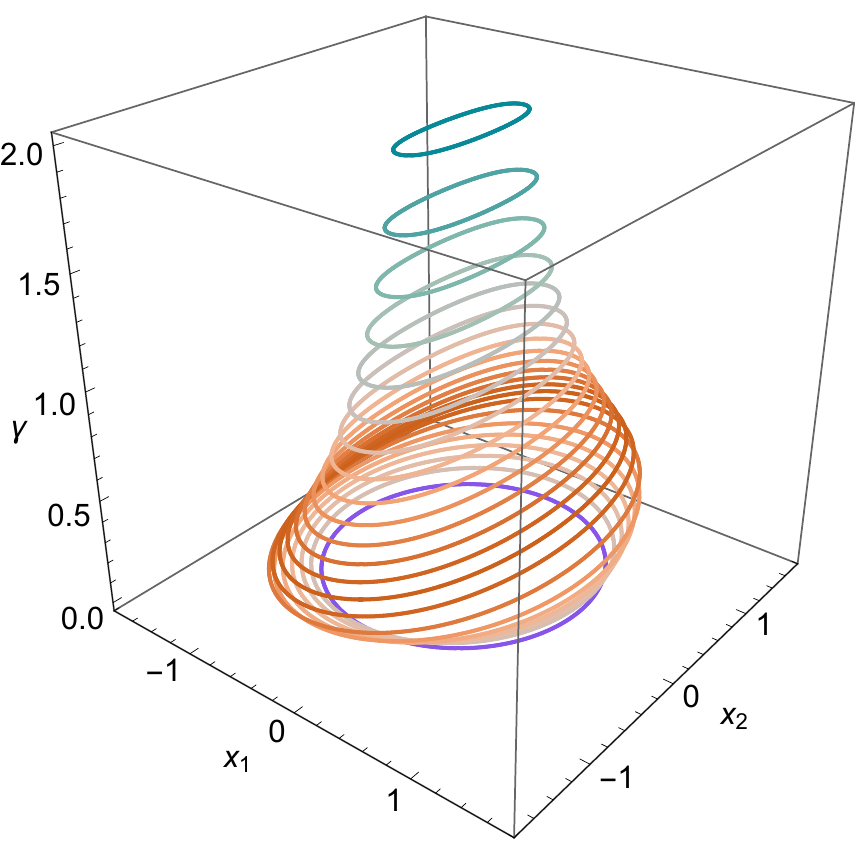}
  \caption{The same worldline instantons as in the third plot of
    Figure~\ref{fig:onedim}, but stacked, with the $z$-coordinate
    given by the parameter $\gamma$. This presentation makes it easier
    to correlate the instantons' change in shape with the parameter.}
  \label{fig:st3d}
\end{figure}

\begin{figure}
  \includegraphics{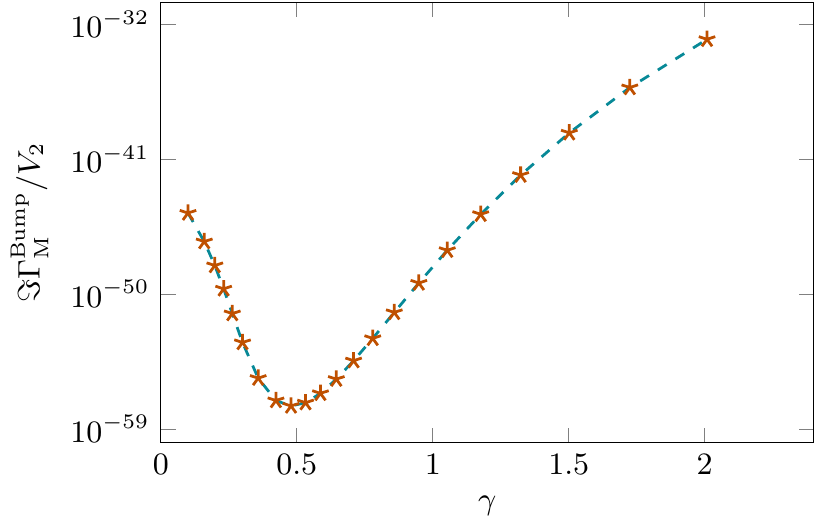}
  \caption{Imaginary part of the Minkowskian effective action for the
    spacetime bump profile
    $\vec{E} = E \cosh^{-2}(\omega t)\cosh^{-2}(k z) \vec{e}_z$ with
    $k = 3\omega$ and $E=0.033m^2/q$. Here (and in the following cases)
    there are no analytical results to compare with, so we just add
    the connecting dashed lines as a guide to the eye.}
  \label{fig:stPP}
\end{figure}

Let us now apply the method outlined above to some background fields.
The strategy in all cases is to start with a limit that is reasonably
close to a static, homogeneous field and perform pseudo-arclength
continuation to map the solution space for a chosen parameter
range. In all figures depicting worldline instantons we color the
homogeneous limit (i.e.\ a circular instanton) purple, and all further
instantons proportional to the change in action (blue for a decrease,
red for an increase, so blue means more, red less pair production). In
all figures that show the full effective action we choose
$E=0.033 m^2/q$ for the reference field strength. This is simply the
value we already used in earlier works, and it does not influence the
quality of the discretization in any way. We also use $N=500$ points
in the discretization, which yields good accuracy while it still takes
less than thirty seconds to obtain the family of instantons in the
cases below, with the exception of the $e$-dipole pulse.

\subsection{Temporal Sauter pulse}

First let us consider simple, one-dimensional inhomogeneities where we
can compare to analytic results. As an example, we choose the
Euclidean four-potential
$iA_3 = \tan(\gamma_\omega x_4)/\gamma_\omega$ describing the
(physical) field $\vec{E} = E \cosh^{-2}(\omega t) \vec{e}_z$ with the
\emph{Keldysh parameter}
$\gamma_\omega=m\omega/qE$~\cite{Keldysh1965}. Since the field does
not depend on any spatial coordinates, we have $N_0=3$ translational
zero modes that need to be held fixed.

The analytical worldline instanton result for this field
is~\cite{Dunne2006a}
\begin{multline}
  \label{eq:tSauterAna}
  \frac{\Im\Gamma_{\up{M}}^{\up{Sauter}^\omega}}{V_3} = \frac{(qE)^{3/2}}{2(2\pi)^3}
  \frac{(1+\gamma_\omega^2)^{5/4}}{\gamma_\omega}\\
  \times\exp\left(
  -\frac{m^2\pi}{qE}\frac{2}{1+\sqrt{1+\gamma_\omega^2}}
  \right).
\end{multline}

In Figure~\ref{fig:onedim} the first plot shows a family of instantons
in the range of $0 < \gamma_\omega < 3.5$, and the top panel in
Figure~\ref{fig:sautPP} compares the numerical
result~\eqref{eq:discrete} in this parameter range to the analytical
expression~\eqref{eq:tSauterAna}, showing near-perfect agreement.

\subsection{Spatial Sauter pulse}

We can also consider the spatially inhomogeneous profile
$iA_4 = \tanh(\gamma_k x_3)/\gamma_k$ describing the (physical) field
$\vec{E} = E \cosh^{-2}(k z) \vec{e}_z$ with (the spatial analog of)
the Keldysh parameter $\gamma_k=mk/qE$. The analytical result is
related to~\eqref{eq:tSauterAna} by
$\gamma_\omega \to i\gamma_k$~\cite{Dunne2006a},
\begin{multline}
  \label{eq:xSauterAna}
  \frac{\Im\Gamma_{\up{M}}^{\up{Sauter}^k}}{V_t V_2} = \frac{(qE)^{3/2}}{2(2\pi)^3}
  \frac{(1-\gamma_k^2)^{5/4}}{\gamma_k}\\
  \times\exp\left(
  -\frac{m^2\pi}{qE}\frac{2}{1+\sqrt{1-\gamma_k^2}}
  \right),
\end{multline}
where the instanton is now confined in $x_3$-direction and we obtain a
``temporal volume factor'' $V_t$ instead. The worldline instantons in
this field for the range $0 < \gamma_k < 1$ are depicted in the middle
of Figure~\ref{fig:onedim}, and the comparison of the numeric result
and the analytic expression~\eqref{eq:xSauterAna} in the bottom panel
of Figure~\ref{fig:sautPP}.

\subsection{Space-time Sauter pulse}

As a simple example of a both space- and time-dependent background we
choose the product of the preceding profiles with
$\gamma := \gamma_\omega = \gamma_k/3$, i.e.\
$iA_3 = \cosh^{-2}(3\gamma x_3)\tan(\gamma x_4)/\gamma$. The resulting
worldline instantons in the range $0 < \gamma < 2.5$ are shown in the
bottom plot of Figure~\ref{fig:onedim} and the resulting pair
production rate in Figure~\ref{fig:stPP}. With the chosen ratio
$\gamma_k/\gamma_\omega=3$ the spatial inhomogeneity dominates for
small $\gamma$, giving larger instantons, increased action and lower
pair production, while above $\gamma \approx 1$ the time dependence
takes over and produces smaller instantons, reduced action and
increased pair production.

\subsection{Multidimensional instantons}

\begin{figure}
  \centering
  \includegraphics[width=\linewidth]{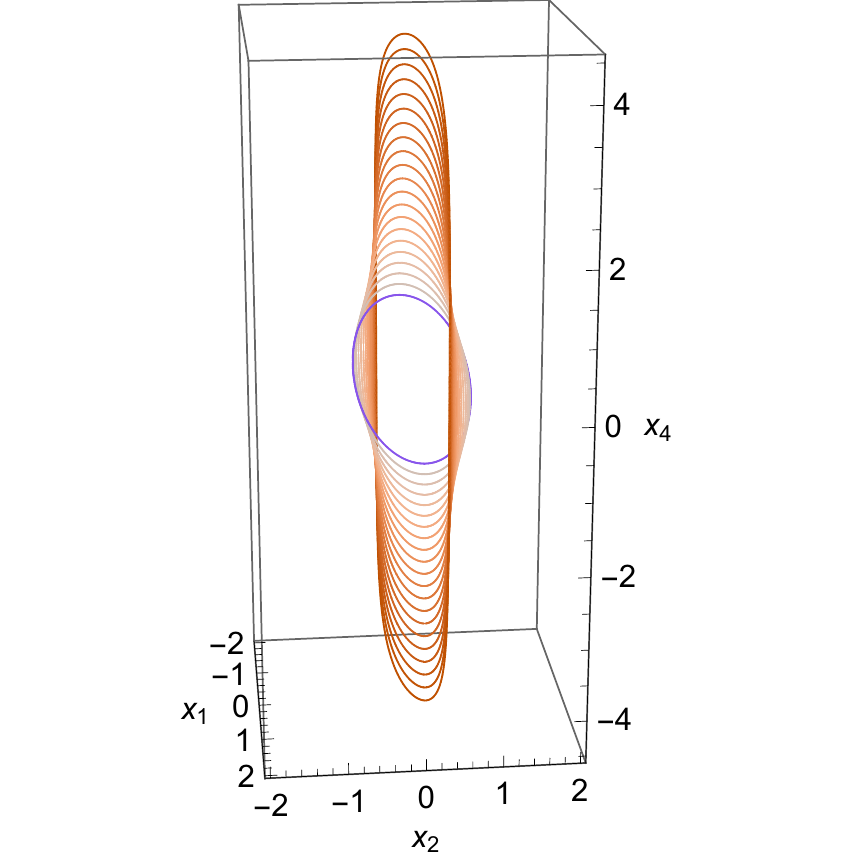}
  \caption{Worldline instantons for the four-potential~\eqref{eq:dunne3d}
    from~\cite{Dunne2006}. As before, stronger
    inhomogeneity stretches the instantons (in a more complicated way
    than for the one dimensional fields) and increases the action.}
  \label{fig:dunne3d}
\end{figure}

\begin{figure}
  \includegraphics{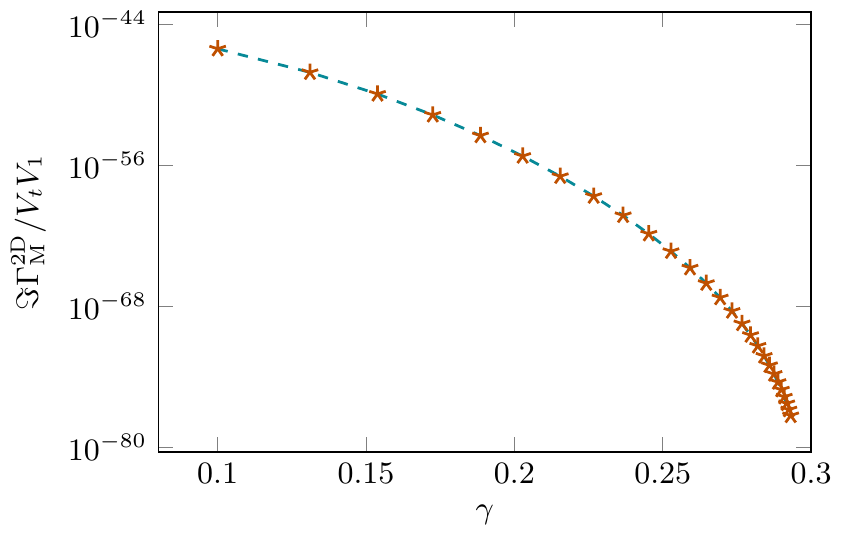}
  \caption{Imaginary part of the effective action for the
    multidimensional field from~\cite{Dunne2006} for
    $E=0.033m^2/q$. The dashed, connecting line is again a
      visual aid only.}
  \label{fig:dunnePP}
\end{figure}

\begin{figure}
  \centering
  \includegraphics{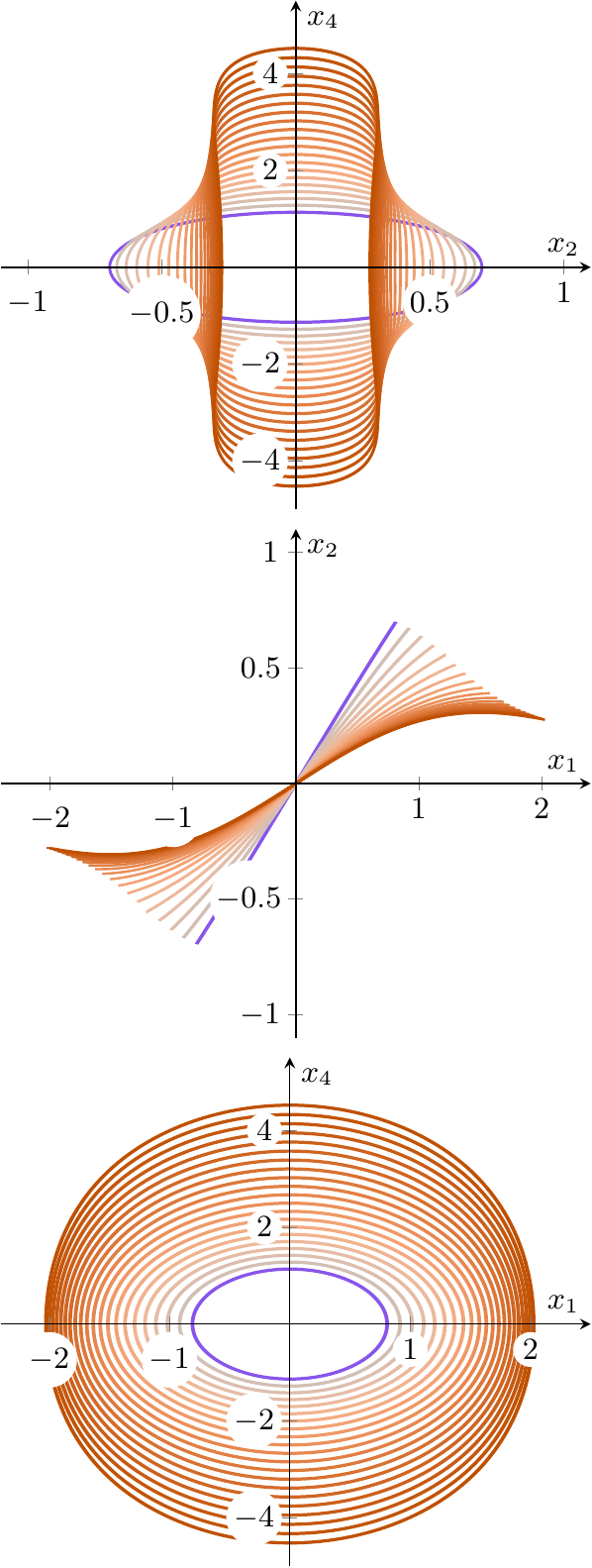}
  \caption{The same worldline instantons as in
    Figure~\ref{fig:dunne3d}, projected onto the coordinate planes.}
  \label{fig:dunneProj}
\end{figure}

In~\cite{Dunne2006} multidimensional instantons were found for
background fields that depend on multiple spatial coordinates using
the shooting method. We can obtain instantons for these fields using
discretization as well. Consider the potential
\begin{equation}
  \label{eq:dunne3d}
iA_4=\frac{1}{\sqrt{2}k}\frac{\tanh(kx_1+kx_2)}{1+(kx_1)^2 + 10(kx_2)^2}
\end{equation}
from Figure~1 in~\cite{Dunne2006} (with the factor of
$1/\sqrt{2}$ added so the peak strength is $1$). This yields three
dimensional instantons (in $x_1$-, $x_2$- and
$x_4$-direction). Figure~\ref{fig:dunne3d} depicts a family of instantons
in a three dimensional plot, while Figure~\ref{fig:dunneProj} shows all
two dimensional projections of the same trajectories. The resulting
pair production rate is given in Figure~\ref{fig:dunnePP}.

\subsection{Plane wave plus electric field}
\begin{figure}
  \centering
  \includegraphics[width=\linewidth]{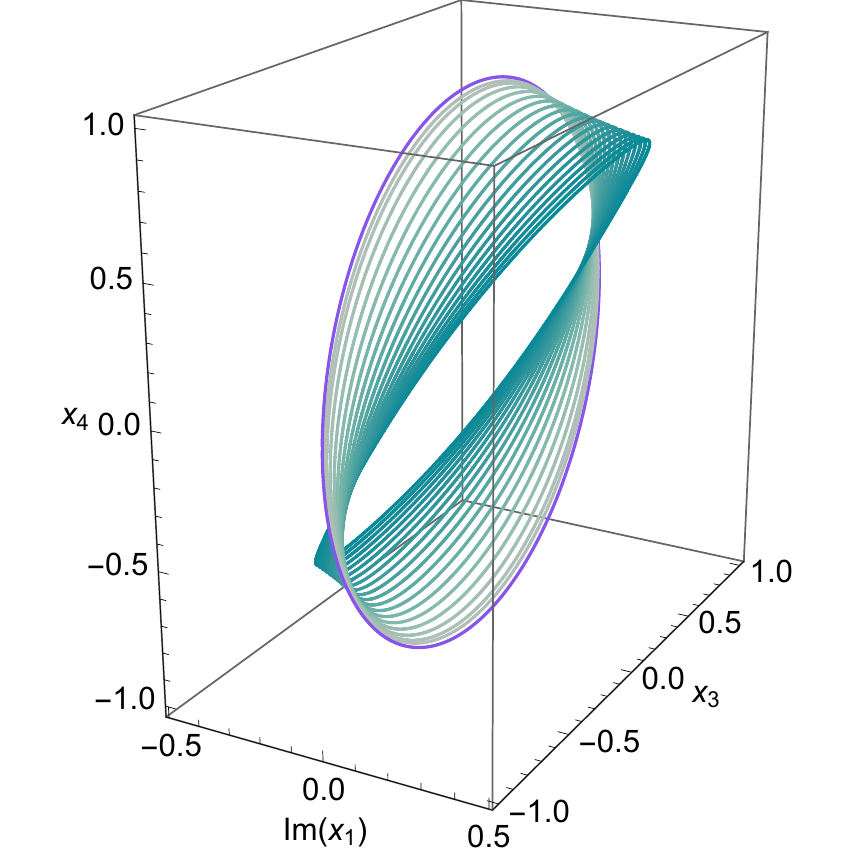}
  \caption{Worldline instantons for the superposition of a static,
    homogeneous field and a weak, propagating plane wave. The ratio of
    the plane wave amplitude and the strong field is $10^{-2}$. The
    $x_1$-component of the trajectories is purely imaginary, hence the
    imaginary part of $x_1$ on the first axis.}
  \label{fig:pw3d}
\end{figure}

\begin{figure}
  \includegraphics{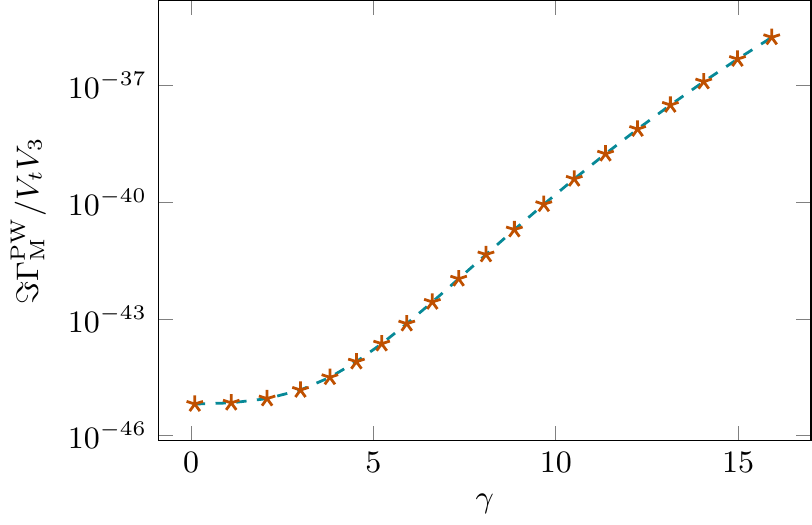}
  \caption{Pair production rate for a constant field
    ($E_{\up{Strong}}=0.033m^2/q$) with superimposed plane wave
    ($E_{\up{Weak}}=10^{-2}E_{\up{Strong}}$). The temporal volume
    factor $V_t$ arises from the number of instantons, one per
    oscillation of the wave at a fixed spatial point. The dashed line
    is added as a guide to the eye.}
  \label{fig:pwPP}
\end{figure}

\begin{figure}
  \centering
  \includegraphics{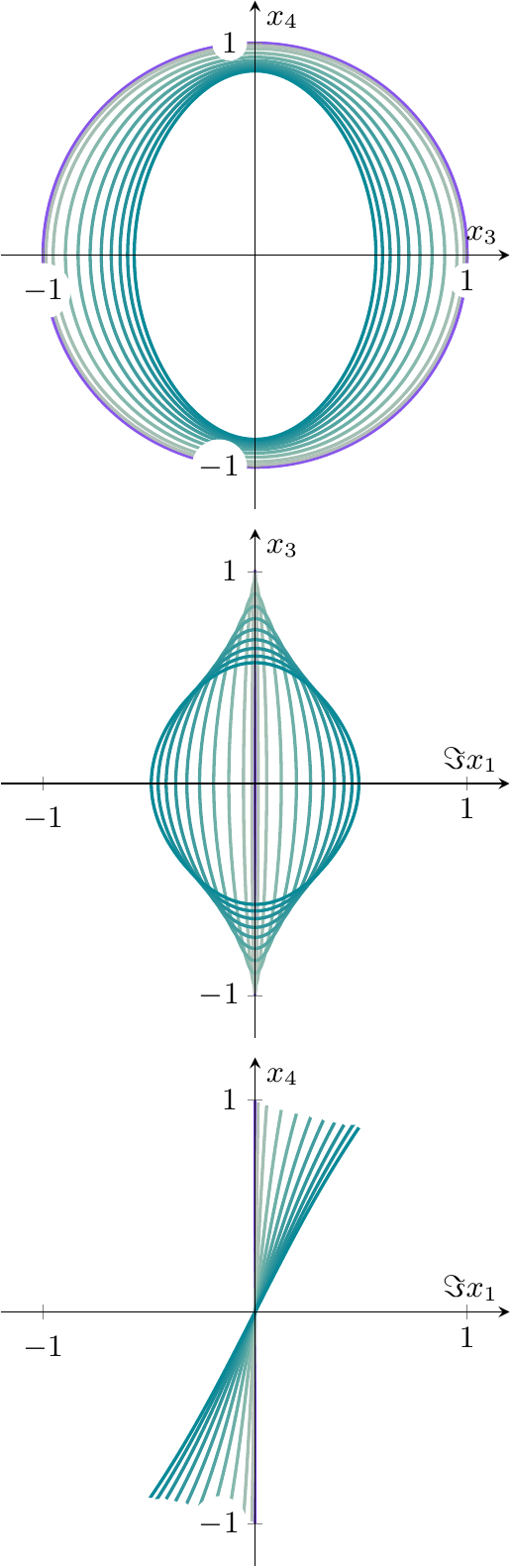}
  \caption{The same worldline instantons as in
    Figure~\ref{fig:pw3d}, projected onto the coordinate planes.}
  \label{fig:pwProj}
\end{figure}

In~\cite{Torgrimsson2017b} we already applied the discrete worldline
instanton method to calculate the pair creation rate for the
superposition of a weak propagating plane wave and a constant field, a
variant of \emph{dynamically assisted} pair
production~\cite{Schutzhold2008}. Different pulse shapes have been
considered for the weak field before~\cite{Linder2015}, however a
plane wave is special in that it cannot produce pairs on its own, so
the process is fully nonperturbative for all frequencies. In the case
of parallel polarization (the plane wave and the constant field point
in the same direction, but perpendicular to the propagation direction)
this combination can be represented by the four-potential
\begin{equation}
  \label{eq:pwPot}
  iA_4 = x_3, \quad iA_3 = 
  i \frac{\varepsilon}{\gamma} \sin\left(\gamma (x_1 - i x_4)\right).
\end{equation}
The method can handle the perpendicularly polarized case just as well,
however that leads to four-dimensional instantons that are cumbersome
to visualize.

In contrast to the examples considered before, the
field~\eqref{eq:pwPot} leads to complex instantons, in particular
purely real $x_3(u)$, $x_4(u)$ and purely imaginary $x_1(u)$. A family
of instantons is shown in Figures~\ref{fig:pw3d} and~\ref{fig:pwProj},
while the full pair production rate is given in Figure~\ref{fig:pwPP}.

\begin{figure}
  \includegraphics{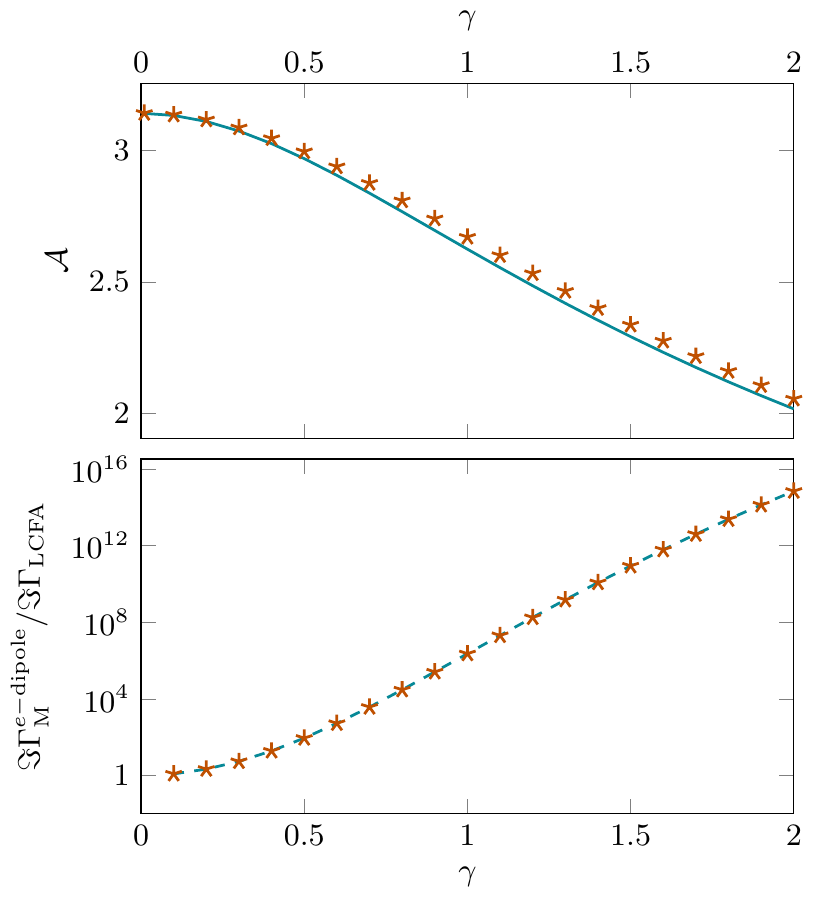}
  \caption{Top: Instanton action for the Gaussian \emph{e}-dipole
    \eqref{eq:edipoleE} (markers) compared to the action for a
    homogeneous field with Gaussian time dependence (line). Bottom:
    Ratio of the effective action and the locally constant field
    approximation for $E = 0.033m^2/q$, with a dashed connecting line
    as a visual aid.}
  \label{fig:eDipoleActPP}
\end{figure}

\subsection{E-dipole pulse}

An especially interesting, highly non-trivial example is that of an
\emph{e-dipole pulse}. It is a solution to Maxwell's equations in
vacuum that represents a localized pulse of finite
energy~\cite{Gonoskov2012}. It saturates the theoretical upper bound
of peak field strength for given laser power~\cite{Bassett1986} and is
thus in a sense the optimal (and at the same time physically viable)
configuration to study pair creation~\cite{Gonoskov2013}. Its name
stems from the structural similarity to dipole radiation, however it
does not suffer from the strong singularities at the origin for a
simple non-stationary dipole.

The electromagnetic field of the \emph{e}-dipole pulse can be given in
terms of a driving function $g$ using the vector
$\vec{Z}$~\cite{Gonoskov2013}
\begin{align}
  \label{eq:edipoleE}
  \begin{split}
    \vec{Z} &= \vec{e}_z \frac{d}{\abs{\vec{r}}}\Big(g(t + \abs{\vec{r}}) - g(t - \abs{\vec{r}})\Big),\\
    \vec{E} &= - \nabla\times\left(\nabla\times\vec{Z}\right), \quad \vec{B} = - \nabla\times\dot{\vec{Z}}.
  \end{split}
\end{align}
We choose the function
\begin{equation}
  g(t) = \frac{t}{4\omega^2}e^{-\omega^2 t^2}+\frac{\sqrt{\pi}}{8\omega^3}(1+2 \omega^2t^2)\erf(\omega t)
\end{equation}
and the virtual dipole moment $d = 3E/4$, so that at the origin
\begin{equation}
  \vec{E} \approx E e^{-\omega^2t^2} \vec{e}_z.
\end{equation}

We cannot immediately apply the instanton approach to this field since
it is not given in terms of a four-potential. It is however possible
to obtain an expression for the potential in coordinate gauge
$A(x)\cdot x = 0$ from the field tensor~\cite{Shifman1980},
\begin{equation}
  A^\up{M}_\mu(x) = -\int\limits_0^1\!\diff{\alpha}\ F^\up{M}_{\mu\nu}(\alpha x)\, \alpha x^\nu.
\end{equation}
For the field~\eqref{eq:edipoleE} this gives a lengthy expression,
which can now be used to obtain worldline instantons.

Figure~\ref{fig:eDipoleActPP} shows the result. The top plot compares
the instanton action for the \emph{e}-dipole pulse to the action for a
field with Gaussian time depencence only.  Due to the additional
spatial inhomogeneity in the \emph{e}-dipole field the action is
slightly larger (and thus pair production slightly lower) than for the
purely time dependent pulse. We can also compare the full imaginary
part of the effective action with the locally constant field
approximation (in the bottom plot of Figure~\ref{fig:eDipoleActPP}),
which can be calculated using the saddle point method for $E$ below
the critical field strength, giving
\begin{equation}
  \Im \Gamma_{\up{LCFA}}^{e\up{-dipole}} \approx \frac{5\sqrt{5}}{2(2\pi)^3 \gamma^4}
  \exp\left(-\pi\frac{m^2}{qE}\right).
\end{equation}

As expected, the worldline instanton result tends to the locally
constant field approximation for small values of $\gamma$, while it is
exponentially larger for higher $\gamma$. For the parameters
considered in~\cite{Gonoskov2013} the adiabaticity is very small, with
$\gamma < 10^{-3}$, so the locally constant field approximation is
accurate. For high frequency pulses however, the pair production rate
is higher than the constant field estimate.

\begin{figure}
  \includegraphics{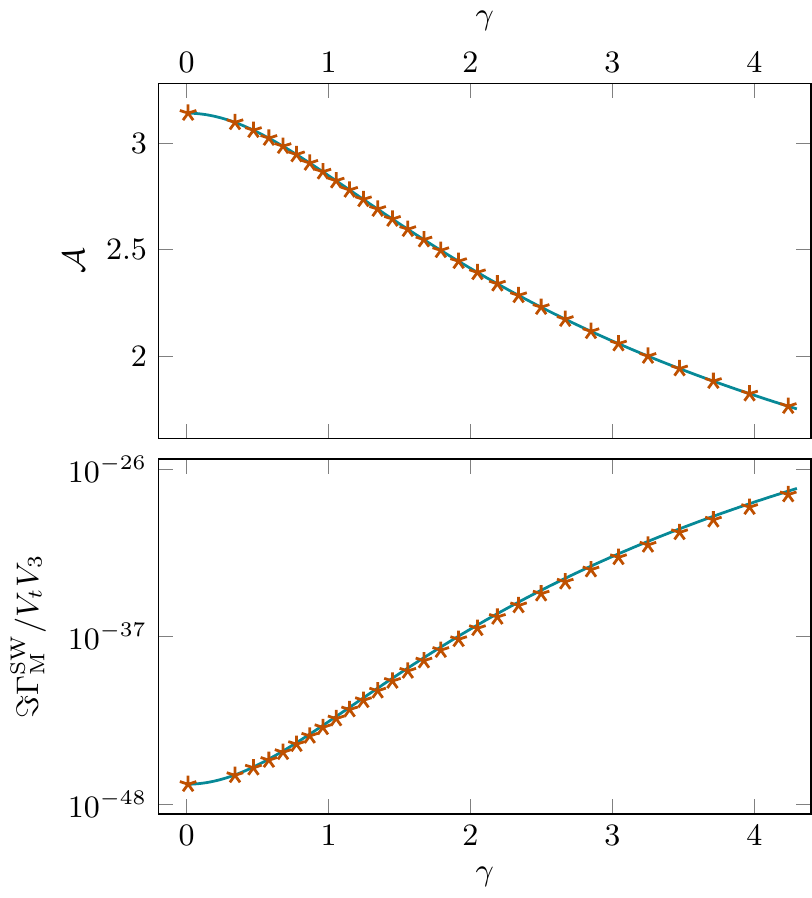}
  \caption{Top: Instanton action for a transversally polarized
    standing wave (markers) compared to just the oscillating time
    dependent field (line). Bottom: The same comparison for the
    imaginary part of the effective action with $E=0.033m^2/q$
    including the prefactor. The transversal inhomogeneity does not
    change the exponent at all, but has a small effect on the
    prefactor.}
  \label{fig:standingWave}
\end{figure}

\subsection{Transversal standing wave}

Let us now briefly consider a purely transversal inhomogeneity. Two
counterpropagating laser beams create a standing wave pattern,
i.e. $\vec{E} = \cos(\omega t)\cos(k x)\vec{e}_z$ with
$k=\omega$. In~\cite{Lv2018} the authors find that omitting the
spatial inhomogeneity leads to qualitatively incorrect results in the
high frequency regime. In \cite{Aleksandrov2017b} strong deviations in
the momentum spectrum have been found in the homogeneous approximation
as well.

In the semiclassical regime and for the total integrated rate however
we can now check that approximating the standing wave by an
oscillating homogeneous field works well. It is easy to see that the
transversal inhomogeneity does not change the instanton and thus the
action~\cite{Linder2015}, but the effect on the prefactor is not as
obvious. Calculating the full effective action using the discrete
instantons shows that while the prefactor does change, the difference
from the homogeneous result is small and barely visible, see
Figure~\ref{fig:standingWave}. Note, however, that the momentum
spectrum could still display noticeable differences between the
standing wave and the purely time dependent field.

\subsection{Constant electric and magnetic fields}
\begin{figure}
  \includegraphics{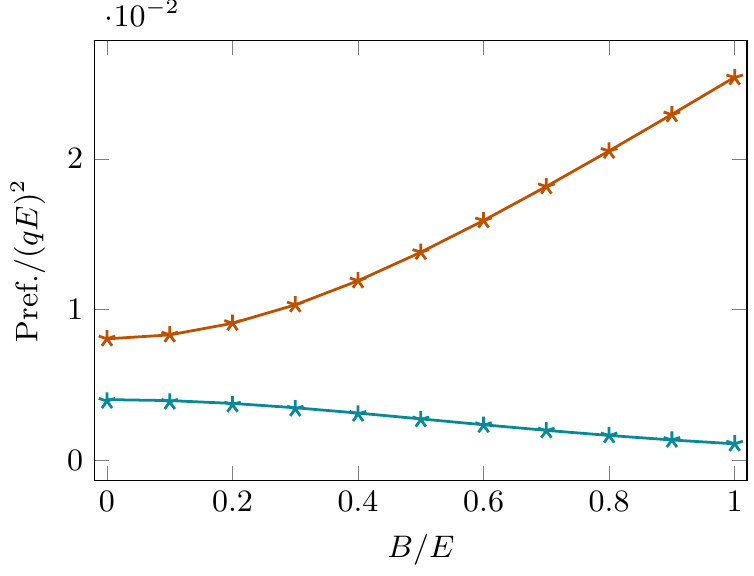}
  \caption{Prefactor divided by $(qE)^2$ for the sum of a constant
    electric field and a magnetic field with varying $B/E$. The
    markers show the discrete instanton result, the lines the analytic
    expressions~\eqref{eq:magAna}. The results for scalar QED are
    blue, the results for spinor QED red.}
  \label{fig:magPrefs}
\end{figure}

In all examples up to now, the spin factor had only a small impact,
apart from the trivial factor of two in the pair production
probability. Let us finally consider a simple example where there is a
large, qualitative difference between scalar and spinor QED, a
parallel superposition of constant electric and magnetic fields of
strength $E$ and $B$ respectively.

The (first term of) the effective action for this combination is given
by (see e.g. \cite{Kim2006} and references therein)
\begin{align}
  \label{eq:magAna}
  \begin{split}
    \Gamma_\mathrm{Scalar} &\approx \frac{(qE)^2}{(2\pi)^3} \pi \frac{B}{E} \csch\left(\pi\frac{B}{E}\right)\exp\left(-\pi\frac{m^2}{qE}\right), \\
    \Gamma_\mathrm{Spinor} &\approx \frac{2(qE)^2}{(2\pi)^3} \pi \frac{B}{E} \coth\left(\pi\frac{B}{E}\right)\exp\left(-\pi\frac{m^2}{qE}\right).
  \end{split}
\end{align}
Figure~\ref{fig:magPrefs} depicts the prefactors of these expressions,
so the $B/E$ dependence, together with the discrete instanton result,
showing perfect agreement.

\section{Summary and conclusion}
We have introduced a new approach to numerically implement the
worldline instanton method for electron-positron pair creation. We use
a discretization scheme that turns the infinite-dimensional path
integral into a finite dimensional integration that we can then
perform using Laplace's method. Crucially, this also means that the
fluctuation prefactor is simply given by a finite dimensional
determinant that can be computed without the great care that is needed
for a properly normalized treatment of the functional determinant.

After having implemented the necessary root finding and continuation
steps outlined in sections~\ref{sec:discretization},
\ref{sec:prefactor} and \ref{sec:continuation}, full pair production
results for arbitrary background fields can be obtained in
minutes. Section~\ref{sec:applications} gives a (by no means
exhaustive) sample of such applications.

Although we used a frequency or inhomogeneity scale as the
continuation parameter in all examples, we could have also chosen a
different field parameter like the polarization direction or the
ellipticity of the field, or even an entirely synthetic parameter to
slowly transition to an especially complicated field configuration.

In this paper we have only considered cases for which there is one
dominant instanton, which is continuously connected to a circular one
in the constant field limit. It would be interesting for future
studies to consider cases where there are more than one instanton, and
where some of them might have a nontrivial topology.

\acknowledgments{We thank Holger Gies and Christian Kohlf\"urst for
  interesting discussions. G.~T. acknowledges support from the
  Alexander von Humboldt foundation.}

\bibliography{library}

\begin{thebibliography}{54}%
\makeatletter
\providecommand \@ifxundefined [1]{%
 \@ifx{#1\undefined}
}%
\providecommand \@ifnum [1]{%
 \ifnum #1\expandafter \@firstoftwo
 \else \expandafter \@secondoftwo
 \fi
}%
\providecommand \@ifx [1]{%
 \ifx #1\expandafter \@firstoftwo
 \else \expandafter \@secondoftwo
 \fi
}%
\providecommand \natexlab [1]{#1}%
\providecommand \enquote  [1]{``#1''}%
\providecommand \bibnamefont  [1]{#1}%
\providecommand \bibfnamefont [1]{#1}%
\providecommand \citenamefont [1]{#1}%
\providecommand \href@noop [0]{\@secondoftwo}%
\providecommand \href [0]{\begingroup \@sanitize@url \@href}%
\providecommand \@href[1]{\@@startlink{#1}\@@href}%
\providecommand \@@href[1]{\endgroup#1\@@endlink}%
\providecommand \@sanitize@url [0]{\catcode `\\12\catcode `\$12\catcode
  `\&12\catcode `\#12\catcode `\^12\catcode `\_12\catcode `\%12\relax}%
\providecommand \@@startlink[1]{}%
\providecommand \@@endlink[0]{}%
\providecommand \url  [0]{\begingroup\@sanitize@url \@url }%
\providecommand \@url [1]{\endgroup\@href {#1}{\urlprefix }}%
\providecommand \urlprefix  [0]{URL }%
\providecommand \Eprint [0]{\href }%
\providecommand \doibase [0]{http://dx.doi.org/}%
\providecommand \selectlanguage [0]{\@gobble}%
\providecommand \bibinfo  [0]{\@secondoftwo}%
\providecommand \bibfield  [0]{\@secondoftwo}%
\providecommand \translation [1]{[#1]}%
\providecommand \BibitemOpen [0]{}%
\providecommand \bibitemStop [0]{}%
\providecommand \bibitemNoStop [0]{.\EOS\space}%
\providecommand \EOS [0]{\spacefactor3000\relax}%
\providecommand \BibitemShut  [1]{\csname bibitem#1\endcsname}%
\let\auto@bib@innerbib\@empty
\bibitem [{\citenamefont {Sauter}(1931)}]{Sauter1931}%
  \BibitemOpen
  \bibfield  {author} {\bibinfo {author} {\bibfnamefont {F.}~\bibnamefont
  {Sauter}},\ }\bibfield  {title} {\enquote {\bibinfo {title} {{Über das
  Verhalten eines Elektrons im homogenen elektrischen Feld nach der
  relativistischen Theorie Diracs}},}\ }\href@noop {} {\bibfield  {journal}
  {\bibinfo  {journal} {Zeitschrift für Physik}\ }\textbf {\bibinfo {volume}
  {69}},\ \bibinfo {pages} {742--764} (\bibinfo {year} {1931})}\BibitemShut
  {NoStop}%
\bibitem [{\citenamefont {Heisenberg}\ and\ \citenamefont
  {Euler}(1936)}]{Heisenberg1936}%
  \BibitemOpen
  \bibfield  {author} {\bibinfo {author} {\bibfnamefont {W.}~\bibnamefont
  {Heisenberg}}\ and\ \bibinfo {author} {\bibfnamefont {H.}~\bibnamefont
  {Euler}},\ }\bibfield  {title} {\enquote {\bibinfo {title} {{Folgerungen aus
  der Diracschen Theorie des Positrons}},}\ }\href@noop {} {\bibfield
  {journal} {\bibinfo  {journal} {Zeitschrift für Physik}\ }\textbf {\bibinfo
  {volume} {98}},\ \bibinfo {pages} {714--732} (\bibinfo {year}
  {1936})}\BibitemShut {NoStop}%
\bibitem [{\citenamefont {Hund}(1941)}]{Hund1941}%
  \BibitemOpen
  \bibfield  {author} {\bibinfo {author} {\bibfnamefont {F.}~\bibnamefont
  {Hund}},\ }\bibfield  {title} {\enquote {\bibinfo {title} {{Materieerzeugung
  im anschaulichen und im gequantelten Wellenbild der Materie}},}\ }\href@noop
  {} {\bibfield  {journal} {\bibinfo  {journal} {Zeitschrift für Physik}\
  }\textbf {\bibinfo {volume} {117}},\ \bibinfo {pages} {1--17} (\bibinfo
  {year} {1941})}\BibitemShut {NoStop}%
\bibitem [{\citenamefont {Schwinger}(1951)}]{Schwinger1951}%
  \BibitemOpen
  \bibfield  {author} {\bibinfo {author} {\bibfnamefont {J.}~\bibnamefont
  {Schwinger}},\ }\bibfield  {title} {\enquote {\bibinfo {title} {On gauge
  invariance and vacuum polarization},}\ }\href {\doibase
  10.1103/PhysRev.82.664} {\bibfield  {journal} {\bibinfo  {journal} {Physical
  Review}\ }\textbf {\bibinfo {volume} {82}},\ \bibinfo {pages} {664} (\bibinfo
  {year} {1951})}\BibitemShut {NoStop}%
\bibitem [{\citenamefont {Cohen}\ and\ \citenamefont
  {McGady}(2008)}]{Cohen2008}%
  \BibitemOpen
  \bibfield  {author} {\bibinfo {author} {\bibfnamefont {T.~D.}\ \bibnamefont
  {Cohen}}\ and\ \bibinfo {author} {\bibfnamefont {D.~A.}\ \bibnamefont
  {McGady}},\ }\bibfield  {title} {\enquote {\bibinfo {title} {The {Schwinger}
  mechanism revisited},}\ }\href {\doibase 10.1103/PhysRevD.78.036008}
  {\bibfield  {journal} {\bibinfo  {journal} {Physical Review D}\ }\textbf
  {\bibinfo {volume} {78}},\ \bibinfo {pages} {036008} (\bibinfo {year}
  {2008})}\BibitemShut {NoStop}%
\bibitem [{Note1()}]{Note1}%
  \BibitemOpen
  \bibinfo {note} {For example the \protect \emph {Extreme Light
  Infrastructure} project \protect \url {https://eli-laser.eu/}}\BibitemShut
  {NoStop}%
\bibitem [{\citenamefont {Kohlfürst}\ and\ \citenamefont
  {Alkofer}(2016)}]{Kohlfurst2016}%
  \BibitemOpen
  \bibfield  {author} {\bibinfo {author} {\bibfnamefont {C.}~\bibnamefont
  {Kohlfürst}}\ and\ \bibinfo {author} {\bibfnamefont {R.}~\bibnamefont
  {Alkofer}},\ }\bibfield  {title} {\enquote {\bibinfo {title} {On the effect
  of time-dependent inhomogeneous magnetic fields in electron–positron pair
  production},}\ }\href {\doibase
  https://doi.org/10.1016/j.physletb.2016.03.027} {\bibfield  {journal}
  {\bibinfo  {journal} {Physics Letters B}\ }\textbf {\bibinfo {volume}
  {756}},\ \bibinfo {pages} {371 -- 375} (\bibinfo {year} {2016})}\BibitemShut
  {NoStop}%
\bibitem [{\citenamefont {Aleksandrov}\ \emph {et~al.}(2017)\citenamefont
  {Aleksandrov}, \citenamefont {Plunien},\ and\ \citenamefont
  {Shabaev}}]{Aleksandrov2017b}%
  \BibitemOpen
  \bibfield  {author} {\bibinfo {author} {\bibfnamefont {I.~A.}\ \bibnamefont
  {Aleksandrov}}, \bibinfo {author} {\bibfnamefont {G.}~\bibnamefont
  {Plunien}}, \ and\ \bibinfo {author} {\bibfnamefont {V.~M.}\ \bibnamefont
  {Shabaev}},\ }\bibfield  {title} {\enquote {\bibinfo {title} {Momentum
  distribution of particles created in space-time-dependent colliding laser
  pulses},}\ }\href {\doibase 10.1103/PhysRevD.96.076006} {\bibfield  {journal}
  {\bibinfo  {journal} {Physical Review D}\ }\textbf {\bibinfo {volume} {96}},\
  \bibinfo {pages} {076006} (\bibinfo {year} {2017})}\BibitemShut {NoStop}%
\bibitem [{\citenamefont {Kohlf\"urst}\ and\ \citenamefont
  {Alkofer}(2018)}]{Kohlfurst2018}%
  \BibitemOpen
  \bibfield  {author} {\bibinfo {author} {\bibfnamefont {C.}~\bibnamefont
  {Kohlf\"urst}}\ and\ \bibinfo {author} {\bibfnamefont {R.}~\bibnamefont
  {Alkofer}},\ }\bibfield  {title} {\enquote {\bibinfo {title} {Ponderomotive
  effects in multiphoton pair production},}\ }\href@noop {} {\bibfield
  {journal} {\bibinfo  {journal} {Phys. Rev. D}\ }\textbf {\bibinfo {volume}
  {97}},\ \bibinfo {pages} {036026} (\bibinfo {year} {2018})}\BibitemShut
  {NoStop}%
\bibitem [{\citenamefont {Lv}\ \emph {et~al.}(2018)\citenamefont {Lv},
  \citenamefont {Dong}, \citenamefont {Li}, \citenamefont {Sheng},
  \citenamefont {Su},\ and\ \citenamefont {Grobe}}]{Lv2018}%
  \BibitemOpen
  \bibfield  {author} {\bibinfo {author} {\bibfnamefont {Q.~Z.}\ \bibnamefont
  {Lv}}, \bibinfo {author} {\bibfnamefont {S.}~\bibnamefont {Dong}}, \bibinfo
  {author} {\bibfnamefont {Y.~T.}\ \bibnamefont {Li}}, \bibinfo {author}
  {\bibfnamefont {Z.~M.}\ \bibnamefont {Sheng}}, \bibinfo {author}
  {\bibfnamefont {Q.}~\bibnamefont {Su}}, \ and\ \bibinfo {author}
  {\bibfnamefont {R.}~\bibnamefont {Grobe}},\ }\bibfield  {title} {\enquote
  {\bibinfo {title} {Role of the spatial inhomogeneity on the laser-induced
  vacuum decay},}\ }\href {\doibase 10.1103/physreva.97.022515} {\bibfield
  {journal} {\bibinfo  {journal} {Physical Review A}\ }\textbf {\bibinfo
  {volume} {97}},\ \bibinfo {pages} {022515} (\bibinfo {year}
  {2018})}\BibitemShut {NoStop}%
\bibitem [{\citenamefont {Aleksandrov}\ \emph {et~al.}()\citenamefont
  {Aleksandrov}, \citenamefont {Plunien},\ and\ \citenamefont
  {Shabaev}}]{Aleksandrov2018}%
  \BibitemOpen
  \bibfield  {author} {\bibinfo {author} {\bibfnamefont {I.~A.}\ \bibnamefont
  {Aleksandrov}}, \bibinfo {author} {\bibfnamefont {G.}~\bibnamefont
  {Plunien}}, \ and\ \bibinfo {author} {\bibfnamefont {V.~M.}\ \bibnamefont
  {Shabaev}},\ }\bibfield  {title} {\enquote {\bibinfo {title} {Dynamically
  assisted {Schwinger} effect beyond the spatially-uniform-field
  approximation},}\ }\href@noop {} {\ }\Eprint {http://arxiv.org/abs/1612.05909
  [hep-ph]} {arXiv:1612.05909 [hep-ph]} \BibitemShut {NoStop}%
\bibitem [{\citenamefont {Kohlf{\"u}rst}(2018)}]{Kohlfurst2018b}%
  \BibitemOpen
  \bibfield  {author} {\bibinfo {author} {\bibfnamefont {C.}~\bibnamefont
  {Kohlf{\"u}rst}},\ }\bibfield  {title} {\enquote {\bibinfo {title}
  {Phase-space analysis of the {Schwinger} effect in inhomogeneous
  electromagnetic fields},}\ }\href@noop {} {\bibfield  {journal} {\bibinfo
  {journal} {The European Physical Journal Plus}\ }\textbf {\bibinfo {volume}
  {133}},\ \bibinfo {pages} {191} (\bibinfo {year} {2018})}\BibitemShut
  {NoStop}%
\bibitem [{\citenamefont {Fock}(1937)}]{Fock1937}%
  \BibitemOpen
  \bibfield  {author} {\bibinfo {author} {\bibfnamefont {V.}~\bibnamefont
  {Fock}},\ }\bibfield  {title} {\enquote {\bibinfo {title} {{Die Eigenzeit in
  der klassischen und in der Quantenmechanik}},}\ }\href@noop {} {\bibfield
  {journal} {\bibinfo  {journal} {Physikalische Zeitschrift der Sowjetunion}\
  }\textbf {\bibinfo {volume} {12}},\ \bibinfo {pages} {404} (\bibinfo {year}
  {1937})}\BibitemShut {NoStop}%
\bibitem [{\citenamefont {Nambu}(1950)}]{Nambu1950}%
  \BibitemOpen
  \bibfield  {author} {\bibinfo {author} {\bibfnamefont {Y.}~\bibnamefont
  {Nambu}},\ }\bibfield  {title} {\enquote {\bibinfo {title} {The use of the
  proper time in quantum electrodynamics {I}},}\ }\href {\doibase
  10.1143/PTP.5.82} {\bibfield  {journal} {\bibinfo  {journal} {Progress of
  Theoretical Physics}\ }\textbf {\bibinfo {volume} {5}},\ \bibinfo {pages}
  {82} (\bibinfo {year} {1950})}\BibitemShut {NoStop}%
\bibitem [{\citenamefont {Feynman}(1950)}]{Feynman1950}%
  \BibitemOpen
  \bibfield  {author} {\bibinfo {author} {\bibfnamefont {R.}~\bibnamefont
  {Feynman}},\ }\bibfield  {title} {\enquote {\bibinfo {title} {Mathematical
  formulation of the quantum theory of electromagnetic interaction},}\ }\href
  {\doibase 10.1103/PhysRev.80.440} {\bibfield  {journal} {\bibinfo  {journal}
  {Physical Review}\ }\textbf {\bibinfo {volume} {80}},\ \bibinfo {pages} {440}
  (\bibinfo {year} {1950})}\BibitemShut {NoStop}%
\bibitem [{\citenamefont {Feynman}(1951)}]{Feynman1951}%
  \BibitemOpen
  \bibfield  {author} {\bibinfo {author} {\bibfnamefont {R.}~\bibnamefont
  {Feynman}},\ }\bibfield  {title} {\enquote {\bibinfo {title} {An operator
  calculus having applications in quantum electrodynamics},}\ }\href {\doibase
  10.1103/PhysRev.84.108} {\bibfield  {journal} {\bibinfo  {journal} {Physical
  Review}\ }\textbf {\bibinfo {volume} {84}},\ \bibinfo {pages} {108} (\bibinfo
  {year} {1951})}\BibitemShut {NoStop}%
\bibitem [{\citenamefont {Gies}\ and\ \citenamefont
  {Langfeld}(2001)}]{Gies2001}%
  \BibitemOpen
  \bibfield  {author} {\bibinfo {author} {\bibfnamefont {H.}~\bibnamefont
  {Gies}}\ and\ \bibinfo {author} {\bibfnamefont {K.}~\bibnamefont
  {Langfeld}},\ }\bibfield  {title} {\enquote {\bibinfo {title} {Quantum
  diffusion of magnetic fields in a numerical worldline approach},}\
  }\href@noop {} {\bibfield  {journal} {\bibinfo  {journal} {Nuclear Physics
  B}\ }\textbf {\bibinfo {volume} {613}},\ \bibinfo {pages} {353 -- 365}
  (\bibinfo {year} {2001})}\BibitemShut {NoStop}%
\bibitem [{\citenamefont {Gies}\ \emph {et~al.}(2003)\citenamefont {Gies},
  \citenamefont {Langfeld},\ and\ \citenamefont {Moyaerts}}]{Gies2003}%
  \BibitemOpen
  \bibfield  {author} {\bibinfo {author} {\bibfnamefont {H.}~\bibnamefont
  {Gies}}, \bibinfo {author} {\bibfnamefont {K.}~\bibnamefont {Langfeld}}, \
  and\ \bibinfo {author} {\bibfnamefont {L.}~\bibnamefont {Moyaerts}},\
  }\bibfield  {title} {\enquote {\bibinfo {title} {Casimir effect on the
  worldline},}\ }\href@noop {} {\bibfield  {journal} {\bibinfo  {journal}
  {Journal of High Energy Physics}\ }\textbf {\bibinfo {volume} {06}},\
  \bibinfo {pages} {018} (\bibinfo {year} {2003})}\BibitemShut {NoStop}%
\bibitem [{\citenamefont {Gies}\ and\ \citenamefont
  {Klingmüller}(2005)}]{Gies2005}%
  \BibitemOpen
  \bibfield  {author} {\bibinfo {author} {\bibfnamefont {H.}~\bibnamefont
  {Gies}}\ and\ \bibinfo {author} {\bibfnamefont {K.}~\bibnamefont
  {Klingmüller}},\ }\bibfield  {title} {\enquote {\bibinfo {title} {Pair
  production in inhomogeneous fields},}\ }\href {\doibase
  10.1103/PhysRevD.72.065001} {\bibfield  {journal} {\bibinfo  {journal}
  {Physical Review D}\ }\textbf {\bibinfo {volume} {72}},\ \bibinfo {pages}
  {065001} (\bibinfo {year} {2005})}\BibitemShut {NoStop}%
\bibitem [{\citenamefont {Gies}\ and\ \citenamefont
  {Roessler}(2011)}]{Gies2011}%
  \BibitemOpen
  \bibfield  {author} {\bibinfo {author} {\bibfnamefont {H.}~\bibnamefont
  {Gies}}\ and\ \bibinfo {author} {\bibfnamefont {L.}~\bibnamefont
  {Roessler}},\ }\bibfield  {title} {\enquote {\bibinfo {title} {Vacuum
  polarization tensor in inhomogeneous magnetic fields},}\ }\href@noop {}
  {\bibfield  {journal} {\bibinfo  {journal} {Phys. Rev. D}\ }\textbf {\bibinfo
  {volume} {84}},\ \bibinfo {pages} {065035} (\bibinfo {year}
  {2011})}\BibitemShut {NoStop}%
\bibitem [{\citenamefont {Affleck}\ \emph {et~al.}(1982)\citenamefont
  {Affleck}, \citenamefont {Alvarez},\ and\ \citenamefont
  {Manton}}]{Affleck1982}%
  \BibitemOpen
  \bibfield  {author} {\bibinfo {author} {\bibfnamefont {I.~K.}\ \bibnamefont
  {Affleck}}, \bibinfo {author} {\bibfnamefont {O.}~\bibnamefont {Alvarez}}, \
  and\ \bibinfo {author} {\bibfnamefont {N.~S.}\ \bibnamefont {Manton}},\
  }\bibfield  {title} {\enquote {\bibinfo {title} {Pair production at strong
  coupling in weak external fields},}\ }\href@noop {} {\bibfield  {journal}
  {\bibinfo  {journal} {Nuclear Physics B}\ }\textbf {\bibinfo {volume}
  {197}},\ \bibinfo {pages} {509} (\bibinfo {year} {1982})}\BibitemShut
  {NoStop}%
\bibitem [{\citenamefont {Dunne}\ and\ \citenamefont
  {Schubert}(2005)}]{Dunne2005}%
  \BibitemOpen
  \bibfield  {author} {\bibinfo {author} {\bibfnamefont {G.~V.}\ \bibnamefont
  {Dunne}}\ and\ \bibinfo {author} {\bibfnamefont {C.}~\bibnamefont
  {Schubert}},\ }\bibfield  {title} {\enquote {\bibinfo {title} {Worldline
  instantons and pair production in inhomogenous fields},}\ }\href {\doibase
  10.1103/PhysRevD.72.105004} {\bibfield  {journal} {\bibinfo  {journal}
  {Physical Review D}\ }\textbf {\bibinfo {volume} {72}},\ \bibinfo {pages}
  {105004} (\bibinfo {year} {2005})}\BibitemShut {NoStop}%
\bibitem [{\citenamefont {Dunne}\ \emph {et~al.}(2006)\citenamefont {Dunne},
  \citenamefont {Wang}, \citenamefont {Gies},\ and\ \citenamefont
  {Schubert}}]{Dunne2006a}%
  \BibitemOpen
  \bibfield  {author} {\bibinfo {author} {\bibfnamefont {G.~V.}\ \bibnamefont
  {Dunne}}, \bibinfo {author} {\bibfnamefont {Q.}~\bibnamefont {Wang}},
  \bibinfo {author} {\bibfnamefont {H.}~\bibnamefont {Gies}}, \ and\ \bibinfo
  {author} {\bibfnamefont {C.}~\bibnamefont {Schubert}},\ }\bibfield  {title}
  {\enquote {\bibinfo {title} {Worldline instantons and the fluctuation
  prefactor},}\ }\href {\doibase 10.1103/PhysRevD.73.065028} {\bibfield
  {journal} {\bibinfo  {journal} {Physical Review D}\ }\textbf {\bibinfo
  {volume} {73}},\ \bibinfo {pages} {065028} (\bibinfo {year}
  {2006})}\BibitemShut {NoStop}%
\bibitem [{\citenamefont {Ilderton}\ \emph {et~al.}(2015)\citenamefont
  {Ilderton}, \citenamefont {Torgrimsson},\ and\ \citenamefont
  {Wårdh}}]{Ilderton2015}%
  \BibitemOpen
  \bibfield  {author} {\bibinfo {author} {\bibfnamefont {A.}~\bibnamefont
  {Ilderton}}, \bibinfo {author} {\bibfnamefont {G.}~\bibnamefont
  {Torgrimsson}}, \ and\ \bibinfo {author} {\bibfnamefont {J.}~\bibnamefont
  {Wårdh}},\ }\bibfield  {title} {\enquote {\bibinfo {title} {Nonperturbative
  pair production in interpolating fields},}\ }\href@noop {} {\bibfield
  {journal} {\bibinfo  {journal} {Physical Review D}\ }\textbf {\bibinfo
  {volume} {92}},\ \bibinfo {pages} {065001} (\bibinfo {year}
  {2015})}\BibitemShut {NoStop}%
\bibitem [{\citenamefont {Schützhold}\ \emph {et~al.}(2008)\citenamefont
  {Schützhold}, \citenamefont {Gies},\ and\ \citenamefont
  {Dunne}}]{Schutzhold2008}%
  \BibitemOpen
  \bibfield  {author} {\bibinfo {author} {\bibfnamefont {R.}~\bibnamefont
  {Schützhold}}, \bibinfo {author} {\bibfnamefont {H.}~\bibnamefont {Gies}}, \
  and\ \bibinfo {author} {\bibfnamefont {G.~V.}\ \bibnamefont {Dunne}},\
  }\bibfield  {title} {\enquote {\bibinfo {title} {Dynamically assisted
  {Schwinger} mechanism},}\ }\href@noop {} {\bibfield  {journal} {\bibinfo
  {journal} {Physical Review Letters}\ }\textbf {\bibinfo {volume} {101}},\
  \bibinfo {pages} {130404} (\bibinfo {year} {2008})}\BibitemShut {NoStop}%
\bibitem [{\citenamefont {Gies}\ and\ \citenamefont
  {Torgrimsson}(2016)}]{Gies2016}%
  \BibitemOpen
  \bibfield  {author} {\bibinfo {author} {\bibfnamefont {H.}~\bibnamefont
  {Gies}}\ and\ \bibinfo {author} {\bibfnamefont {G.}~\bibnamefont
  {Torgrimsson}},\ }\bibfield  {title} {\enquote {\bibinfo {title} {Critical
  schwinger pair production},}\ }\href@noop {} {\bibfield  {journal} {\bibinfo
  {journal} {Physical Review Letters}\ }\textbf {\bibinfo {volume} {116}},\
  \bibinfo {pages} {090406} (\bibinfo {year} {2016})}\BibitemShut {NoStop}%
\bibitem [{\citenamefont {Schneider}\ and\ \citenamefont
  {Schützhold}(2016)}]{Schneider2016}%
  \BibitemOpen
  \bibfield  {author} {\bibinfo {author} {\bibfnamefont {C.}~\bibnamefont
  {Schneider}}\ and\ \bibinfo {author} {\bibfnamefont {R.}~\bibnamefont
  {Schützhold}},\ }\bibfield  {title} {\enquote {\bibinfo {title} {Dynamically
  assisted sauter-schwinger effect in inhomogeneous electric fields},}\
  }\href@noop {} {\bibfield  {journal} {\bibinfo  {journal} {Journal of High
  Energy Physics}\ }\textbf {\bibinfo {volume} {2}},\ \bibinfo {pages} {1--16}
  (\bibinfo {year} {2016})}\BibitemShut {NoStop}%
\bibitem [{\citenamefont {Dunne}\ and\ \citenamefont {Wang}(2006)}]{Dunne2006}%
  \BibitemOpen
  \bibfield  {author} {\bibinfo {author} {\bibfnamefont {G.~V.}\ \bibnamefont
  {Dunne}}\ and\ \bibinfo {author} {\bibfnamefont {Q.}~\bibnamefont {Wang}},\
  }\bibfield  {title} {\enquote {\bibinfo {title} {Multidimensional worldline
  instantons},}\ }\href {\doibase 10.1103/PhysRevD.74.065015} {\bibfield
  {journal} {\bibinfo  {journal} {Physical Review D}\ }\textbf {\bibinfo
  {volume} {74}},\ \bibinfo {pages} {065015} (\bibinfo {year}
  {2006})}\BibitemShut {NoStop}%
\bibitem [{\citenamefont {Schubert}(2001)}]{Schubert2001}%
  \BibitemOpen
  \bibfield  {author} {\bibinfo {author} {\bibfnamefont {C.}~\bibnamefont
  {Schubert}},\ }\bibfield  {title} {\enquote {\bibinfo {title} {Perturbative
  quantum field theory in the string-inspired formalism},}\ }\href@noop {}
  {\bibfield  {journal} {\bibinfo  {journal} {Physics Reports}\ }\textbf
  {\bibinfo {volume} {355}},\ \bibinfo {pages} {73} (\bibinfo {year}
  {2001})}\BibitemShut {NoStop}%
\bibitem [{\citenamefont {Schubert}(2012)}]{Schubert2012}%
  \BibitemOpen
  \bibfield  {author} {\bibinfo {author} {\bibfnamefont {C.}~\bibnamefont
  {Schubert}},\ }\bibfield  {title} {\enquote {\bibinfo {title} {Lectures on
  the worldline formalism},}\ }\href@noop {} {\bibfield  {journal} {\bibinfo
  {journal} {School on Spinning Particles in Quantum Field Theory: Worldline
  Formalism, Higher Spins and Conformal Geometry}\ } (\bibinfo {year}
  {2012})}\BibitemShut {NoStop}%
\bibitem [{\citenamefont {Gies}\ and\ \citenamefont
  {H{ä}mmerling}(2005)}]{Gies2005a}%
  \BibitemOpen
  \bibfield  {author} {\bibinfo {author} {\bibfnamefont {H.}~\bibnamefont
  {Gies}}\ and\ \bibinfo {author} {\bibfnamefont {J.}~\bibnamefont
  {H{ä}mmerling}},\ }\bibfield  {title} {\enquote {\bibinfo {title} {Geometry
  of spin-field coupling on the worldline},}\ }\href {\doibase
  10.1103/PhysRevD.72.065018} {\bibfield  {journal} {\bibinfo  {journal}
  {Physical Review D}\ }\textbf {\bibinfo {volume} {72}},\ \bibinfo {pages}
  {1--20} (\bibinfo {year} {2005})}\BibitemShut {NoStop}%
\bibitem [{\citenamefont {Feynman}(1948)}]{feynman1948}%
  \BibitemOpen
  \bibfield  {author} {\bibinfo {author} {\bibfnamefont {R.}~\bibnamefont
  {Feynman}},\ }\bibfield  {title} {\enquote {\bibinfo {title} {Space-time
  approach to non-relativistic quantum mechanics},}\ }\href {\doibase
  10.1103/RevModPhys.20.367} {\bibfield  {journal} {\bibinfo  {journal}
  {Reviews of Modern Physics}\ }\textbf {\bibinfo {volume} {20}},\ \bibinfo
  {pages} {367} (\bibinfo {year} {1948})}\BibitemShut {NoStop}%
\bibitem [{\citenamefont {Rabello}\ and\ \citenamefont
  {Farina}(1995)}]{Rabello1995}%
  \BibitemOpen
  \bibfield  {author} {\bibinfo {author} {\bibfnamefont {S.~J.}\ \bibnamefont
  {Rabello}}\ and\ \bibinfo {author} {\bibfnamefont {C.}~\bibnamefont
  {Farina}},\ }\bibfield  {title} {\enquote {\bibinfo {title} {Gauge invariance
  and the path integral},}\ }\href@noop {} {\bibfield  {journal} {\bibinfo
  {journal} {Physical Review A}\ }\textbf {\bibinfo {volume} {51}},\ \bibinfo
  {pages} {2614–2615} (\bibinfo {year} {1995})}\BibitemShut {NoStop}%
\bibitem [{\citenamefont {Stone}(2000)}]{Stone2000}%
  \BibitemOpen
  \bibfield  {author} {\bibinfo {author} {\bibfnamefont {M.}~\bibnamefont
  {Stone}},\ }\href {\doibase 10.1007/978-1-4612-0507-4} {\emph {\bibinfo
  {title} {The Physics of Quantum Fields}}},\ Graduate Texts in Contemporary
  Physics\ (\bibinfo  {publisher} {Springer New York},\ \bibinfo {year}
  {2000})\BibitemShut {NoStop}%
\bibitem [{\citenamefont {Gaveau}\ \emph {et~al.}(2004)\citenamefont {Gaveau},
  \citenamefont {Mihóková}, \citenamefont {Roncadelli},\ and\ \citenamefont
  {Schulman}}]{Gaveau2004}%
  \BibitemOpen
  \bibfield  {author} {\bibinfo {author} {\bibfnamefont {B.}~\bibnamefont
  {Gaveau}}, \bibinfo {author} {\bibfnamefont {E.}~\bibnamefont {Mihóková}},
  \bibinfo {author} {\bibfnamefont {M.}~\bibnamefont {Roncadelli}}, \ and\
  \bibinfo {author} {\bibfnamefont {L.~S.}\ \bibnamefont {Schulman}},\
  }\bibfield  {title} {\enquote {\bibinfo {title} {Path integral in a magnetic
  field using the {Trotter} product formula},}\ }\href@noop {} {\bibfield
  {journal} {\bibinfo  {journal} {American Journal of Physics}\ }\textbf
  {\bibinfo {volume} {72}},\ \bibinfo {pages} {385–388} (\bibinfo {year}
  {2004})}\BibitemShut {NoStop}%
\bibitem [{\citenamefont {Schulman}(2005)}]{Schulman2005}%
  \BibitemOpen
  \bibfield  {author} {\bibinfo {author} {\bibfnamefont {L.~S.}\ \bibnamefont
  {Schulman}},\ }\href@noop {} {\emph {\bibinfo {title} {Techniques and
  Applications of Path Integration}}}\ (\bibinfo  {publisher} {Dover
  Publications, New York},\ \bibinfo {year} {2005})\BibitemShut {NoStop}%
\bibitem [{\citenamefont {D’Hoker}\ and\ \citenamefont
  {Gagné}(1996)}]{DHoker1996}%
  \BibitemOpen
  \bibfield  {author} {\bibinfo {author} {\bibfnamefont {E.}~\bibnamefont
  {D’Hoker}}\ and\ \bibinfo {author} {\bibfnamefont {D.~G.}\ \bibnamefont
  {Gagné}},\ }\bibfield  {title} {\enquote {\bibinfo {title} {Worldline path
  integrals for fermions with scalar, pseudoscalar and vector couplings},}\
  }\href {\doibase 10.1016/0550-3213(96)00125-3} {\bibfield  {journal}
  {\bibinfo  {journal} {Nuclear Physics B}\ }\textbf {\bibinfo {volume}
  {467}},\ \bibinfo {pages} {272} (\bibinfo {year} {1996})}\BibitemShut
  {NoStop}%
\bibitem [{\citenamefont {Gies}\ and\ \citenamefont
  {Langfeld}(2002)}]{Gies2002}%
  \BibitemOpen
  \bibfield  {author} {\bibinfo {author} {\bibfnamefont {H.}~\bibnamefont
  {Gies}}\ and\ \bibinfo {author} {\bibfnamefont {K.}~\bibnamefont
  {Langfeld}},\ }\bibfield  {title} {\enquote {\bibinfo {title} {Loops and loop
  clouds - a numerical approach to the worldline formalism},}\ }\href@noop {}
  {\bibfield  {journal} {\bibinfo  {journal} {International Journal of Modern
  Physics A}\ }\textbf {\bibinfo {volume} {17}},\ \bibinfo {pages} {966--976}
  (\bibinfo {year} {2002})}\BibitemShut {NoStop}%
\bibitem [{\citenamefont {Faddeev}\ and\ \citenamefont
  {Popov}(1967)}]{Faddeev1967}%
  \BibitemOpen
  \bibfield  {author} {\bibinfo {author} {\bibfnamefont {L.~D.}\ \bibnamefont
  {Faddeev}}\ and\ \bibinfo {author} {\bibfnamefont {V.~N.}\ \bibnamefont
  {Popov}},\ }\bibfield  {title} {\enquote {\bibinfo {title} {Feynman diagrams
  for the {Yang-Mills} field},}\ }\href {\doibase 10.1016/0370-2693(67)90067-6}
  {\bibfield  {journal} {\bibinfo  {journal} {Physics Letters}\ }\textbf
  {\bibinfo {volume} {25B}},\ \bibinfo {pages} {29} (\bibinfo {year}
  {1967})}\BibitemShut {NoStop}%
\bibitem [{\citenamefont {Gordon}\ and\ \citenamefont
  {Semenoff}(2015)}]{Gordon2015}%
  \BibitemOpen
  \bibfield  {author} {\bibinfo {author} {\bibfnamefont {J.}~\bibnamefont
  {Gordon}}\ and\ \bibinfo {author} {\bibfnamefont {G.~W.}\ \bibnamefont
  {Semenoff}},\ }\bibfield  {title} {\enquote {\bibinfo {title} {World-line
  instantons and the {Schwinger} effect as a {Wentzel-Kramers-Brillouin} exact
  path integral},}\ }\href {\doibase 10.1063/1.4908556} {\bibfield  {journal}
  {\bibinfo  {journal} {Journal of Mathematical Physics}\ }\textbf {\bibinfo
  {volume} {56}},\ \bibinfo {pages} {022111} (\bibinfo {year}
  {2015})}\BibitemShut {NoStop}%
\bibitem [{\citenamefont {Gordon}\ and\ \citenamefont
  {Semenoff}(2016)}]{Gordon2016}%
  \BibitemOpen
  \bibfield  {author} {\bibinfo {author} {\bibfnamefont {J.}~\bibnamefont
  {Gordon}}\ and\ \bibinfo {author} {\bibfnamefont {G.~W.}\ \bibnamefont
  {Semenoff}},\ }\href@noop {} {\enquote {\bibinfo {title} {{Schwinger} pair
  production: Explicit localization of the world-line instanton},}\ } (\bibinfo
  {year} {2016}),\ \Eprint {http://arxiv.org/abs/1612.05909} {arXiv:1612.05909}
  \BibitemShut {NoStop}%
\bibitem [{\citenamefont {Zinn-Justin}(1996)}]{ZinnJustin1996}%
  \BibitemOpen
  \bibfield  {author} {\bibinfo {author} {\bibfnamefont {J.}~\bibnamefont
  {Zinn-Justin}},\ }\href@noop {} {\emph {\bibinfo {title} {Quantum field
  theory and critical phenomena}}},\ International Series of Monographs on
  Physics\ (\bibinfo  {publisher} {Clarendon Press},\ \bibinfo {year}
  {1996})\BibitemShut {NoStop}%
\bibitem [{\citenamefont {Allgower}\ and\ \citenamefont
  {Georg}(2003)}]{Allgower2003}%
  \BibitemOpen
  \bibfield  {author} {\bibinfo {author} {\bibfnamefont {E.~L.}\ \bibnamefont
  {Allgower}}\ and\ \bibinfo {author} {\bibfnamefont {K.}~\bibnamefont
  {Georg}},\ }\href {\doibase 10.1137/1.9780898719154} {\emph {\bibinfo {title}
  {Introduction to Numerical Continuation Methods}}}\ (\bibinfo  {publisher}
  {Society for Industrial and Applied Mathematics},\ \bibinfo {year}
  {2003})\BibitemShut {NoStop}%
\bibitem [{\citenamefont {Rheinboldt}(2000)}]{Rheinboldt2000}%
  \BibitemOpen
  \bibfield  {author} {\bibinfo {author} {\bibfnamefont {W.~C.}\ \bibnamefont
  {Rheinboldt}},\ }\bibfield  {title} {\enquote {\bibinfo {title} {Numerical
  continuation methods: a perspective},}\ }\href@noop {} {\bibfield  {journal}
  {\bibinfo  {journal} {Journal of Computational and Applied Mathematics}\
  }\textbf {\bibinfo {volume} {124}},\ \bibinfo {pages} {229} (\bibinfo {year}
  {2000})}\BibitemShut {NoStop}%
\bibitem [{\citenamefont {Gould}\ and\ \citenamefont
  {Rajantie}(2017)}]{Gould2017}%
  \BibitemOpen
  \bibfield  {author} {\bibinfo {author} {\bibfnamefont {O.}~\bibnamefont
  {Gould}}\ and\ \bibinfo {author} {\bibfnamefont {A.}~\bibnamefont
  {Rajantie}},\ }\bibfield  {title} {\enquote {\bibinfo {title} {Thermal
  {Schwinger} pair production at arbitrary coupling},}\ }\href {\doibase
  10.1103/physrevd.96.076002} {\bibfield  {journal} {\bibinfo  {journal}
  {Physical Review D}\ }\textbf {\bibinfo {volume} {96}},\ \bibinfo {pages}
  {076002} (\bibinfo {year} {2017})}\BibitemShut {NoStop}%
\bibitem [{\citenamefont {Davidenko}(1953)}]{Davidenko1953}%
  \BibitemOpen
  \bibfield  {author} {\bibinfo {author} {\bibfnamefont {D.~F.}\ \bibnamefont
  {Davidenko}},\ }\bibfield  {title} {\enquote {\bibinfo {title} {{On a new
  method of numerically integrating a system of nonlinear equations
  (Russian)}},}\ }\href@noop {} {\bibfield  {journal} {\bibinfo  {journal}
  {Doklady Akademii Nauk SSSR}\ }\textbf {\bibinfo {volume} {88}},\ \bibinfo
  {pages} {601} (\bibinfo {year} {1953})}\BibitemShut {NoStop}%
\bibitem [{\citenamefont {Keldysh}(1965)}]{Keldysh1965}%
  \BibitemOpen
  \bibfield  {author} {\bibinfo {author} {\bibfnamefont {L.~V.}\ \bibnamefont
  {Keldysh}},\ }\bibfield  {title} {\enquote {\bibinfo {title} {Ionization in
  the field of a strong electromagnetic wave},}\ }\href@noop {} {\bibfield
  {journal} {\bibinfo  {journal} {Journal of Experimental and Theoretical
  Physics}\ }\textbf {\bibinfo {volume} {20}},\ \bibinfo {pages} {1307}
  (\bibinfo {year} {1965})}\BibitemShut {NoStop}%
\bibitem [{\citenamefont {Torgrimsson}\ \emph {et~al.}(2018)\citenamefont
  {Torgrimsson}, \citenamefont {Schneider},\ and\ \citenamefont
  {Schützhold}}]{Torgrimsson2017b}%
  \BibitemOpen
  \bibfield  {author} {\bibinfo {author} {\bibfnamefont {G.}~\bibnamefont
  {Torgrimsson}}, \bibinfo {author} {\bibfnamefont {C.}~\bibnamefont
  {Schneider}}, \ and\ \bibinfo {author} {\bibfnamefont {R.}~\bibnamefont
  {Schützhold}},\ }\bibfield  {title} {\enquote {\bibinfo {title}
  {{Sauter-Schwinger pair creation dynamically assisted by a plane wave}},}\
  }\href@noop {} {\bibfield  {journal} {\bibinfo  {journal} {Physical Review
  D}\ }\textbf {\bibinfo {volume} {97}},\ \bibinfo {pages} {096004} (\bibinfo
  {year} {2018})}\BibitemShut {NoStop}%
\bibitem [{\citenamefont {Linder}\ \emph {et~al.}(2015)\citenamefont {Linder},
  \citenamefont {Schneider}, \citenamefont {Sicking}, \citenamefont {Szpak},\
  and\ \citenamefont {Schützhold}}]{Linder2015}%
  \BibitemOpen
  \bibfield  {author} {\bibinfo {author} {\bibfnamefont {M.~F.}\ \bibnamefont
  {Linder}}, \bibinfo {author} {\bibfnamefont {C.}~\bibnamefont {Schneider}},
  \bibinfo {author} {\bibfnamefont {J.}~\bibnamefont {Sicking}}, \bibinfo
  {author} {\bibfnamefont {N.}~\bibnamefont {Szpak}}, \ and\ \bibinfo {author}
  {\bibfnamefont {R.}~\bibnamefont {Schützhold}},\ }\bibfield  {title}
  {\enquote {\bibinfo {title} {Pulse shape dependence in the dynamically
  assisted {Sauter-Schwinger} effect},}\ }\href {\doibase
  10.1103/PhysRevD.92.085009} {\bibfield  {journal} {\bibinfo  {journal}
  {Physical Review D}\ }\textbf {\bibinfo {volume} {92}},\ \bibinfo {pages}
  {085009} (\bibinfo {year} {2015})}\BibitemShut {NoStop}%
\bibitem [{\citenamefont {Gonoskov}\ \emph {et~al.}(2012)\citenamefont
  {Gonoskov}, \citenamefont {Aiello}, \citenamefont {Heugel},\ and\
  \citenamefont {Leuchs}}]{Gonoskov2012}%
  \BibitemOpen
  \bibfield  {author} {\bibinfo {author} {\bibfnamefont {I.}~\bibnamefont
  {Gonoskov}}, \bibinfo {author} {\bibfnamefont {A.}~\bibnamefont {Aiello}},
  \bibinfo {author} {\bibfnamefont {S.}~\bibnamefont {Heugel}}, \ and\ \bibinfo
  {author} {\bibfnamefont {G.}~\bibnamefont {Leuchs}},\ }\bibfield  {title}
  {\enquote {\bibinfo {title} {{Dipole pulse theory: Maximizing the field
  amplitude from $4\pi$ focused laser pulses}},}\ }\href {\doibase
  10.1103/PhysRevA.86.053836} {\bibfield  {journal} {\bibinfo  {journal}
  {Physical Review A}\ }\textbf {\bibinfo {volume} {86}},\ \bibinfo {pages}
  {053836} (\bibinfo {year} {2012})}\BibitemShut {NoStop}%
\bibitem [{\citenamefont {Bassett}(1986)}]{Bassett1986}%
  \BibitemOpen
  \bibfield  {author} {\bibinfo {author} {\bibfnamefont {I.~M.}\ \bibnamefont
  {Bassett}},\ }\bibfield  {title} {\enquote {\bibinfo {title} {Limit to
  concentration by focusing},}\ }\href {\doibase 10.1080/713821943} {\bibfield
  {journal} {\bibinfo  {journal} {Optica Acta: International Journal of
  Optics}\ }\textbf {\bibinfo {volume} {33}},\ \bibinfo {pages} {279} (\bibinfo
  {year} {1986})}\BibitemShut {NoStop}%
\bibitem [{\citenamefont {Gonoskov}\ \emph {et~al.}(2013)\citenamefont
  {Gonoskov}, \citenamefont {Gonoskov}, \citenamefont {Harvey}, \citenamefont
  {Ilderton}, \citenamefont {Kim}, \citenamefont {Marklund}, \citenamefont
  {Mourou},\ and\ \citenamefont {Sergeev}}]{Gonoskov2013}%
  \BibitemOpen
  \bibfield  {author} {\bibinfo {author} {\bibfnamefont {A.}~\bibnamefont
  {Gonoskov}}, \bibinfo {author} {\bibfnamefont {I.}~\bibnamefont {Gonoskov}},
  \bibinfo {author} {\bibfnamefont {C.}~\bibnamefont {Harvey}}, \bibinfo
  {author} {\bibfnamefont {A.}~\bibnamefont {Ilderton}}, \bibinfo {author}
  {\bibfnamefont {A.}~\bibnamefont {Kim}}, \bibinfo {author} {\bibfnamefont
  {M.}~\bibnamefont {Marklund}}, \bibinfo {author} {\bibfnamefont
  {G.}~\bibnamefont {Mourou}}, \ and\ \bibinfo {author} {\bibfnamefont
  {A.}~\bibnamefont {Sergeev}},\ }\bibfield  {title} {\enquote {\bibinfo
  {title} {Probing nonperturbative {QED} with optimally focused laser
  pulses},}\ }\href {\doibase 10.1103/PhysRevLett.111.060404} {\bibfield
  {journal} {\bibinfo  {journal} {Physical Review Letters}\ }\textbf {\bibinfo
  {volume} {111}},\ \bibinfo {pages} {060404} (\bibinfo {year}
  {2013})}\BibitemShut {NoStop}%
\bibitem [{\citenamefont {Shifman}(1980)}]{Shifman1980}%
  \BibitemOpen
  \bibfield  {author} {\bibinfo {author} {\bibfnamefont {M.~A.}\ \bibnamefont
  {Shifman}},\ }\bibfield  {title} {\enquote {\bibinfo {title} {Wilson loop in
  vacuum fields},}\ }\href {\doibase 10.1016/0550-3213(80)90440-x} {\bibfield
  {journal} {\bibinfo  {journal} {Nuclear Physics B}\ }\textbf {\bibinfo
  {volume} {173}},\ \bibinfo {pages} {13–31} (\bibinfo {year}
  {1980})}\BibitemShut {NoStop}%
\bibitem [{\citenamefont {Kim}\ and\ \citenamefont {Page}(2006)}]{Kim2006}%
  \BibitemOpen
  \bibfield  {author} {\bibinfo {author} {\bibfnamefont {S.~P.}\ \bibnamefont
  {Kim}}\ and\ \bibinfo {author} {\bibfnamefont {D.}~\bibnamefont {Page}},\
  }\bibfield  {title} {\enquote {\bibinfo {title} {Schwinger pair production in
  electric and magnetic fields},}\ }\href {\doibase 10.1103/PhysRevD.73.065020}
  {\bibfield  {journal} {\bibinfo  {journal} {Physical Review D}\ }\textbf
  {\bibinfo {volume} {73}},\ \bibinfo {pages} {065020} (\bibinfo {year}
  {2006})}\BibitemShut {NoStop}%
\end{thebibliography}%

\end{document}